\documentclass[11pt,twoside]{article}
\usepackage[utf8]{inputenc}
\usepackage{amsmath,amssymb,latexsym}
\usepackage{amsthm}
\usepackage{a4wide}
\usepackage{fourier}
\usepackage{array}
\usepackage{faktor}
\usepackage{fancyhdr}
\usepackage{graphicx}
\usepackage[svgnames]{xcolor}
\usepackage[labelfont=bf,width=0.8\textwidth,justification=centering]{caption}
\usepackage{tikz}
\usepackage{subfigure}
\usepackage[colorlinks=true,linkcolor=Indigo,citecolor=Indigo]{hyperref}

\newtheorem{theorem}{Theorem}[section]
\newtheorem{proposition}[theorem]{Proposition}
\newtheorem{lemma}[theorem]{Lemma}
\theoremstyle{definition}
\newtheorem{rmk}[theorem]{Remark}

\numberwithin{equation}{section}
\numberwithin{figure}{section}

\title{Two body problem on a sphere in the presence of a uniform magnetic field}
\author{Nataliya A. Balabanova \ \& \ James Montaldi}
\date{May 2021\footnote{Regular and Chaotic Dynamics, to appear 2021}}

\begin{document}
  \maketitle 

\vskip-5mm
\hfill\parbox{3cm}{\centering\textbf{In memory of \\ Alexey V. Borisov}}

  \begin{abstract}
      We investigate the motion of one and two charged non-relativistic particles on a sphere in the presence of a magnetic field of uniform strength. For one particle, the motion is always circular, and determined by a simple relation between the velocity and the radius of motion. For two identical particles, interacting via a cotangent potential, we show there are two families of relative equilibria, called Type I and Type II. The Type I relative equilibria exist for all strengths of the magnetic field, while those of Type II exist only if the field is sufficiently strong. The same is true if the particles are of equal mass but opposite charge.  We also determine the stability of the two families of relative equilibria.


\end{abstract}
  
 \tableofcontents

\section{Introduction}

There have been a number of studies on the dynamics of two charged particles in the plane and in space in the presence of a uniform magnetic field \cite{pinheiro2006interaction, pinheiro2008interaction,escobar2015two}. In each of these, the symmetry group consists of rotations and translations in the plane (that is, $\mathsf{SE}(2)$), together with translations along the magnetic field direction in the spatial setting.

In this paper we consider a similar problem, but with the particles constrained to move on the surface of a sphere, with the magnetic field vector normal to the surface and of constant magnitude. This ensures the system has spherical symmetry; the symmetry group is therefore the group of rotations $\mathsf{SO}(3)$.

Physically, this setup requires there to be a net magnetic charge within the sphere, for example a magnetic monopole at the centre of the sphere. 
The motion of a charged particle in $\mathbb{R}^3$ in the presence of a monopole is a well-studied topic, and goes back to Poincar\'e in 1896 \cite{Poincare1896} who showed that the motion of the particle is constrained to a circular cone.
More recent work can be seen in, for example \cite{McIntosh} and \cite{Zwanziger}. Various generalizations of the Kepler problem to $S^3$, among which is the addition of a spherical analog of the magnetic monopole, are studied in \cite{borisov2005superintegrable}. 

The traditional Lagrangian or Hamiltonian approach to studying systems in the presence of a magnetic field is to introduce a vector potential.  However, on the sphere there is no globally defined vector potential (for topological reasons), so we use the alternative Hamiltonian approach which incorporates the magnetic field into the symplectic structure, similar to the treatment in \cite{pinheiro2006interaction,pinheiro2008interaction} and first introduced by Littlejohn \cite{littlejohn1979guiding}.

\subsubsection*{Reduction and particle motion}
We investigate the  motion of one and two particles on a sphere in the presence of a centrally symmetric magnetic field using a Hamiltonian approach. Similarly to their conterparts with no magnetic vector field, the system with one particle is an integrable one; however that is not the case for two particles.

For a single charged particle, the motion is either stationary or has circles as trajectories, with a simple relation between the velocity of the particle and the radius of the circle, depending on the strength of the magnetic field (Proposition \ref{st: one part}).

For two particles, we adapt the well-known Hamiltonian approach for the similar non-magnetic problem \cite{borisov2004two,borisov2018reduction} to our setting and perform the Poisson reduction with respect to the $\mathsf{SO}(3)$-action, obtaining the reduced Hamiltonian, Hamilton's equations and the Poisson structure. The latter is degenerate and possesses a Casimir function, the second first integral of motion, which involves the angular momentum and the strength of the magnetic field. 

The reduced system is described in terms of the distance $q$ between the particles, its conjugate variable $p$ and the three components of the angular momentum in the body frame $(m_1,m_2,m_3)$, similar to the analogous problem with no magnetic field \cite{borisov2018reduction}. The magnetic field $B$ explicitly enters through the form of the Poisson structure. 

The goal of this paper is to find the relative equilibrium states of the system. Solving the equations explicitly in the general case does not appear to be tractable, due to their complexity. Nonetheless, in Theorem \ref{st: Existence of eq gen case} we establish the general existence of relative equilibria for arbitrary masses and charges of the two particles.

\subsubsection*{Identical particles}

For calculations, we choose the explicit potential  $V(q) = e_1e_2\cot(q)$,  where $e_1$ and $e_2$ are the charges of the particles. This function arises as a fundamental solution of the Laplace-Beltrami equation for the case of an electric field due to a single charged particle  on a three-dimensional sphere $S^3$. It also arises as an instance of the solutions of the generalized Bertrand problem of finding potentials depending only on geodesic distance and with closed orbits  \cite{kozlov1994dynamics}. As physics dictates, the potential is repelling when the signs of the charges coincide and attracting when they differ.

Due to the complexity of the general form of the reduced system of equations, in Section 5 we limit our attention to the (tractable) case where the two particles are identical. 

Solving the equations for identical particles explicitly gives four classes of relative equilibria. Two of these---called Type I relative equilibria---exist for all values of $(q,B)$ with $q\neq\pi/2$, while the other two---those of Type II---exist only in a certain region of the $(q,B)$-plane.
The two Type I relative equilibria are related by an exchange of particles but the two Type II relative equilibria are genuinely different.

Reconstructing motion from the solutions of the reduced system of equations requires  expressing the vector of angular velocity of the solutions and noticing that it has to be parallel to the value of the momentum map, $\Phi$ for that configuration. 

\begin{figure} 
\centering

\subfigure[Obtuse and acute relative equilibrium of Type I.]{%
    \begin{tikzpicture}[scale=1.5]  
        \draw [thick] (0,0) circle (1cm);
            \draw [-latex,dashed] (0,-1) -- (0,1.4);
            \draw [-latex] (0.2,1.15) arc (-30:250: 2mm and 0.5mm);
            \draw (-0.3,1.15) node {$\boldsymbol\omega$};
            \draw [thin] (0,0) -- (.6012, .7991);
            \draw (0,0.2) arc (90:53.1:2mm);
            \filldraw [violet] (.6012, .7991) circle (2pt);
            \draw [thin] (0,0) -- (.3915, -.9202);
            \draw (0,-0.2) arc (-90:-67:2mm);
            \draw (0,-0.25) arc (-90:-63:2mm);
            \filldraw [violet] (.3915, -.9202) circle (2pt);
     \end{tikzpicture}
         \hskip 6mm
    \begin{tikzpicture}[scale=1.5] 
        \draw [thick] (0,0) circle (1cm);
            \draw [-latex,dashed] (0,-1) -- (0,1.4);
            \draw [-latex] (0.2,1.15) arc (-30:250: 2mm and 0.5mm);
            \draw (-0.3,1.15) node {$\boldsymbol\omega$};
            \draw [thin] (0,0) -- (.7993, .6010);
            \draw (0,0.2) arc (90:37:2mm);
            \filldraw [violet] (.7993, .6010) circle (2pt);
            \draw [thin] (0,0) -- (.9204, -.3911);
            \draw (0,-0.2) arc (-90:-23:2mm);
            \draw (0,-0.25) arc (-90:-23:2.5mm);
            \draw (0, -1) node {};
            \filldraw [violet] (.9204, -.3911) circle (2pt);
 \end{tikzpicture}
 } \hskip15mm
\subfigure[Obtuse and acute relative equilibrium of Type II.]{%
    \begin{tikzpicture}[scale=1.5]  
        \draw [thick] (0,0) circle (1cm);
            \draw [-latex] (0.2,1.5) arc (-30:250: 2mm and 0.5mm);
            \draw [-latex,dashed] (0,-1) -- (0,1.4);
              \draw [-latex,dashed] (0,-1) -- (0,1.8);
            \draw [-latex] (0.2,1.15) arc (-30:250: 2mm and 0.5mm);
            \draw (-0.35,1.15) node {$\boldsymbol\omega_1$};
            \draw (-0.35,1.5) node {$\boldsymbol\omega_2$};
            \draw [thin] (0,0) -- (0.866, -0.5);
            \draw (0,-0.2) arc (-90:-30.1:2mm);
           
            \draw [thin] (0,0) -- (-0.866, -0.5);
            \draw (0,-0.2) arc (-90:-150:2mm);
             \filldraw [violet] (-0.866, -0.5) circle (2 pt);
            \filldraw [violet] (0.866, -0.5) circle (2 pt);
     \end{tikzpicture}
         \hskip 6mm
   \begin{tikzpicture}[scale=1.5]  
        \draw [thick] (0,0) circle (1cm);
            \draw [-latex,dashed] (0,-1) -- (0,1.4);
            \draw [-latex,dashed] (0,-1) -- (0,1.8);
            \draw [-latex] (0.2,1.15) arc (-30:250: 2mm and 0.5mm);
             \draw [-latex] (0.2,1.5) arc (-30:250: 2mm and 0.5mm);
            \draw (-0.35,1.15) node {$\boldsymbol\omega_1$};
            \draw (-0.35,1.5) node {$\boldsymbol\omega_2$};
            \draw [thin] (0,0) -- ( -0.5,-0.866);
            \draw (0,-0.2) arc (-90:-60:2mm);
            \filldraw [violet] ( -0.5,-0.866) circle (2 pt);
            \draw [thin] (0,0) -- ( 0.5,-0.866);
            \draw (0,-0.2) arc (-90:-120:2mm);
            \filldraw [violet] (0.5,-0.866) circle (2 pt);
     \end{tikzpicture}
 }
 
 \caption{The two types of relative equilibrium for equal masses and equal charges, with $V(q) = \cot(q)$. (See Figure\,\ref{fig:RE opposite charges} for the case with opposite charges.)  }
 \label{fig:acute and obtuse RE}
\end{figure}
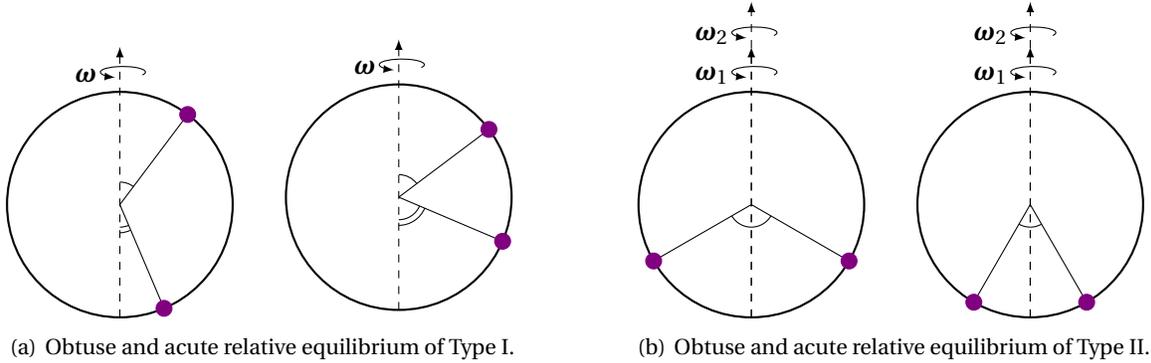
 
Type I relative equilibria exist with both acute and obtuse angles between the bodies (but not a right angle); in both cases, the motions of the particles are \textit{parallel} in the sense that the axis of rotation is located to the side of the two particles, and the angles of the two particles with the axis of rotation are always different.  
 
On the other hand, Type II relative equilibria have  \textit{isosceles} configurations: be they obtuse, acute or right-angled, the axis of rotation is always placed between the bodies, equidistant from both of the particles. They have another interesting property: the same geometric configuration can be occupied by two states with different rates of rotation, and hence different energy levels.

Employing the standard method of investigating linear stability of relative equilibria, we find the regions in the $(q,B)$-plane for which the equilibria in question are stable or unstable.

Type I relative equilibria allow for an analytic solution of the region of stability, described in Section \ref{sec:Type I stability}, where we show that larger inter-particle distance are more likely to be stable.   In contrast, for Type II relative equilibria, we need to employ  numerical methods to differentiate between stability and instability (see Section \ref{sec:Type II stability}).

We finish the analysis of the stabilityby calculating the Hessians of different types of relative equilibrium and plotting them on a simplified version of the energy-momentum diagram. 

In Section 6 we briefly discuss the setting with two particles of equal mass but opposite charge. Indeed, we note that there is a simple transformation of the phase space that takes the case of identical charges to the case of opposite charges, extending a similar observation in \cite{garcia2020attracting}.  Due to the nature of the transformations required for the switch, all the relative equilibria are retained, together with their respective stability properties, but the actual motion of the particles will change.

The case of two identical particles with a cotangent potential has a limiting case when $B\to 0$: the two-body problem with a repelling potential described in \cite{garcia2020attracting}. We draw comparisons to it throughout the paper, and in the first appendix we discuss the limiting case of relative equilibria: the right-angle ones. Type I relative equilibria do not exist for $q= \frac{\pi}{2}$ when $B\ne 0$, and 
Type II relative equilibria are not present for small values of $B$.  It turns out that different right-angled relative equilibria for the gravitational problem appear as limits of the equilibria of Type I when we approach the point $(q,B)=(\frac{\pi}{2},0)$ via different curves $B= B(q)$. 


\section{The setup}
We begin by recalling some basic facts from electrodynamics. 

The force acting on a non-relativistic particle with mass $\mu$ and charge $e$ in the presence of an electric field $\mathbf{E}$ and a magnetic field $\mathbf{B}$ is called the Lorenz force and is given by  $\mathbf{ F} = e(\mathbf{E}  + \mathbf{v}\times\mathbf{B})$, where $\mathbf{v}$ is the velocity of the particle.  In the absence of any other forces acting on the particle, Newton's second law dictates that 
\begin{equation}
    \label{Lorenz}
   \mu\mathbf{a} = e(\mathbf{E}  + \mathbf{v}\times\mathbf{B}), 
\end{equation}
with $\mathbf{a}$ denoting the acceleration of the particle.

Recall that from Maxwell's equations (see, for example, \cite{landau1971classical}), $\nabla \cdot\mathbf{B} = 0$.

Using Hodge duality in $\mathbb{R}^3$, $\mathbf{B}$ can be represented as a two-form rather than a vector field: $\mathbf{B} = \mathbf{B}_3\mathrm{d}x\wedge \mathrm{d}y +\mathbf{B}_2\mathrm{d}z\wedge \mathrm{d}x +\mathbf{B}_1\mathrm{d}y\wedge \mathrm{d}z$. 
The divergence-free nature of the vector field is then rephrased as stating that $\mathbf{B}$ is a closed two-form.

Consider now  two non-relativistic particles in $\mathbb{R}^3$, with respective masses $\mu_1$ and $\mu_2$, charges $e_1$ and $e_2$, position vectors $\mathbf{q}_1$ and $\mathbf{q}_2$ and velocities $\mathbf{v}_1,\,\mathbf{v}_2$, and momenta $\mathbf{p}_1,\,\mathbf{p}_2$ (where $\mathbf{p}_i=\mu_i\mathbf{v}_i$). 

It is well known that their motion is described by the trajectories of a Hamiltonian system on $T^{\ast}\mathbb{R}^3\times T^{\ast}\mathbb{R}^3$, with Hamiltonian 
  \begin{equation}
  \label{Ham}
        \mathcal{H} = \frac{1}{2\mu_1}\left|\mathbf{p}_1\right|^2 +  \frac{1}{2\mu_2}\left|\mathbf{p}_2\right|^2 + V(\mathbf{q}_1,\mathbf{q}_2)
    \end{equation}
    and symplectic form 
    \begin{equation}
    \label{omega}
     \omega =   \left( \begin{array}{c|c}
           \begin{array}{c c}
               e_1\mathfrak{B} & 0 \\
               0 & e_2\mathfrak{B}
           \end{array}& I_6\\
           \hline
           -I_6& 0
        \end{array}\right),
    \end{equation}
where  $\mathfrak{B}$ is the matrix 
\begin{equation} \label{magnetic 2-form}
\mathfrak{B}= \begin{pmatrix}
   0& -B_3&B_2\\
    B_3& 0& -B_1\\
    -B_2&B_1&0
    \end{pmatrix},
\end{equation}
which represents the magnetic 2-form. The function  $V(\mathbf{q}_1,\mathbf{q}_2)$ is the potential energy of the system, borne out of the interaction between the particles and the effect of any pre-existing electric field. 

First, we note that due to closedness of $\mathbf{B}$, $\omega$ is a closed form as well. Together with its obvious non-degeneracy and skew symmetry, this shows \eqref{omega} and \eqref{Ham} form a Hamiltonian system.

To prove that this Hamiltonian system does indeed model the system, we need to demonstrate that the Hamiltonian equations obtained from \eqref{Ham}  are equivalent to Newton's second law, as written for each of the particles:
    
\begin{equation}
\label{Ham eqs}
\left\{    \begin{array}{lcl}
    \dot{\mathbf{q}}_1 &=& \frac{1}{\mu_1}\mathbf{p}_1,\\[4pt]
    \dot{\mathbf{q}}_2 &=& \frac{1}{\mu_2}\mathbf{p}_2,\\[4pt]
    \dot{\mathbf{p}}_1  &=& -V'(\mathbf{q}_1, \mathbf{q}_2)_{\mathbf{q}_1} + \frac{e_1}{\mu_1}\mathfrak{B}|_{\mathbf{q}_1}( \mathbf{p}_1),\\[6pt]
     \dot{\mathbf{p}}_2  &=& -V'(\mathbf{q}_1 ,\mathbf{q}_2)_{\mathbf{q}_2} + \frac{e_2}{\mu_2}\mathfrak{B}|_{\mathbf{q}_2}(\mathbf{p}_2).
    \end{array}\right.
\end{equation}

The first two lines in \eqref{Ham eqs} are tautological expressions, being equivalent to the definitions of $\mathbf{p}_i$. The second two are precisely Newton's second law. Thus, Hamiltonian equations on two copies of the cotangent bundle of $\mathbb{R}^3$ with the symplectic form given above are equivalent to Newton equations for particle motion.

\subsubsection*{Setup for two particles on the plane and in space}

Here we recall briefly the approach to the problem taken by the authors in  \cite{escobar2015two,pinheiro2006interaction,pinheiro2008interaction} and \cite{littlejohn1979guiding}.

The study by Escobar-Ruiz and Turbiner \cite{escobar2015two} for two particles in the plane is based on a Lagrangian approach involving a vector potential for the magnetic field (that is, $\mathbf{A}$ satisfying $\mathbf{B}=\nabla\times\mathbf{A}$, or as differential forms, $\mathbf{B}=\mathrm{d}\mathbf{A}$). 

Littlejohn \cite{littlejohn1979guiding} showed how to avoid the use of a vector potential by incorporating the magnetic field (as a 2-form) into the symplectic form, and then proceed directly with Hamilton's formulation. This was used by him to study the guiding centre problem, and more pertinent for us,  was also used by Pinheiro and MacKay \cite{pinheiro2006interaction,pinheiro2008interaction} in their studies of two particles in the plane and in space.  

For a uniform magnetic field on the sphere, there is no vector potential (the existence of a vector potential implies the magnetic field has mean zero, by Stokes' theorem), so we are obliged to use the Hamiltonian approach described above.

All the authors assume (as do we) that there is no external electric field, so that the potential is a function of the distance between the particles $V(\mathbf{q}_1, \mathbf{q}_2)=V\left(\|\mathbf{q}_1 - \mathbf{q}_2\|\right)$. 
Hamilton's equations for the motion in the plane are still given by \eqref{Ham eqs}.

Particles are placed on the horizontal plane in \cite{escobar2015two} and \cite{pinheiro2006interaction} and in three-dimensional space in \cite{pinheiro2008interaction}. These systems have similarly structured symmetry groups, comprised of translations (in the plane or in space) and rotations in the $x$-$y$-plane.

The specific magnetic field studied in all cases is (as a 2-form),  $\mathbf{B} = B\mathrm{d}x\wedge \mathrm{d}y$, with $B$ constant, which corresponds to a uniform magnetic field parallel to the $z$-axis.

\section{Motion of one particle}

Suppose that one particle of mass $\mu$ and charge $e$ is placed on a unit sphere. 

  We take  $V(\mathbf{q}) = 0$ and the magnetic vector field $\mathbf{B}$ orthogonal to $S^2$ of uniform strength $B$;  thus, it can be easily seen that the system possesses $\mathsf{SO}(3)$-symmetry. 

It can be checked that the restriction of our system to $T^{\ast}S^2$ will be a Hamiltonian system; this is a simple reasoning that we omit here.

Keeping notation as above, we may easily write the symplectic form (on $T^*\mathbb{R}^3$, not restricted to $T^{\ast}S^2$ in this notation) as
\begin{equation}
\label{sympl str one p}
  \omega =  \begin{pmatrix}
  e\mathfrak{B} & I\\-I&0
  \end{pmatrix}
\end{equation}
where $\mathfrak{B}$ is given in Eq.\,\eqref{magnetic 2-form}. The Hamiltonian and the momentum map on $T^*S^2$ are 
\begin{equation}
    \label{Ham one p}
    \mathcal{H} = \frac{1}{2\mu}|\mathbf{p}|^2
\end{equation}
\begin{equation}
    \label{mom map one p}
\Phi: T^{\ast}S^2\to\mathfrak{so}(3)^{\ast}, \  \  \Phi(\mathbf{q},\mathbf{p}) = -eB\mathbf{q} + \mathbf{q}\times\mathbf{p}
\end{equation}

The momentum map $\Phi$ is a map from a  4-dimensional space into a three dimensional one and its fibres $\Phi^{-1}(\eta), \  \eta\in\mathfrak{so}(3)^{\ast}$  are one-dimensional.

Additionally, the stabilizer of any non-zero element in $\mathfrak{so}(3)^{\ast}$ is isomorphic to $\mathsf{SO}(2)$ and has dimension one. Thus, as quotient spaces, all the reduced spaces $\faktor{\Phi^{-1}(\eta)}{\mathsf{SO}(3)_{\eta}}$ are single points, making any motion a relative (or absolute, in case of a stationary particle) equilibrium.

Explicitly writing the formulae for the preimage of an element in $\mathfrak{so}(3)^{\ast}$ and finding the maximal Euclidean distance between any two points therein (keeping in mind that it has to be a circle) gives 

\begin{proposition}
\label{st: one part}
The trajectories of a non-relativistic charged particle on a unit sphere with a unform magnetic field are circles, whose radius $r$ is related to the velocity $\mathbf{v}$ by the relation
\begin{equation}
    r^2 = \frac{\mu^2|\mathbf{v}|^2}{B^2e^2 + \mu^2|\mathbf{v}|^2}
\end{equation}
In terms of angular velocity $\boldsymbol\omega$ this is,
\begin{equation}
  r^2=1-\frac{e^2B^2}{\mu^2|\boldsymbol\omega|^2}
\end{equation}
\end{proposition}

The circle on which the particle moves is the intersection of the sphere with the cone given by Poincar\'e's solution to the motion in $\mathbb{R}^3$ of a charged particle in the presence of a monopole. Indeed, the expression for the momentum map in \eqref{mom map one p} is the restriction to $T^*S^2$ of the one for the action of $\mathsf{SO}(3)$ on $T^*\mathbb{R}^3$ with symplectic form \eqref{sympl str one p}, which is
$$(\mathbf{x},\mathbf{p})\longmapsto -\frac{eB}{|\mathbf{x}|}\mathbf{x}+\mathbf{x}\times\mathbf{p}.$$
That this expression is a conserved quantity for the motion of a charged particle in the presence of a magnetic monopole was shown by Poincar\'e \cite{Poincare1896}.

\section{Motion of two particles} \label{sec:Motion of two particles} 

Now consider the setup with two particles of respective masses $\mu_1$ and $\mu_2$ and charges $e_1$ and $e_2$, located, once again, on a unit sphere.

 As previously, we assume that no external electric fields are present, resulting in $V(\mathbf{q}_1,\mathbf{q}_2) = V(||\mathbf{q}_1 - \mathbf{q}_2||)$, and throughout we assume $V'(q)\neq0$ for all $q\in(0,\pi)$.

Following \cite{borisov2018reduction}, we assume that $\lim\limits_{q\to0^+} V(q) \to\infty$ and $\lim\limits_{q\to \pi^-} V(q) \to-\infty$ (or vice versa, when $V(q)$ is an attracting potential). Thus, the configuration space of the problem is $\mathcal{Q} = S^2\times S^2\setminus \Delta$, where $\Delta$ is the union of the diagonal subset of $S^2\times S^2$ and the subset that contains all the pairs of antipodal points.

The momentum map for the $\mathsf{SO}(3)$ action on the phase space $T^*\mathcal{Q}$ is the sum of those for each particle:
\begin{equation}
\label{mm}
    \Phi(\mathbf{q}_1,\mathbf{q}_2,\mathbf{p}_1,\mathbf{p}_2) = -B(e_1\mathbf{q}_1 +e_2\mathbf{q}_2) +\mathbf{q}_1\times \mathbf{p}_1 +\mathbf{q}_2\times \mathbf{p}_2
\end{equation}

For reduction, we follow along the  lines of parametrization used in \cite{borisov2018reduction,borisov2004two} and \cite{garcia2020attracting}.

Any matrix in $\mathsf{SO}(3)$ can be written using Euler angles in the form $g(\theta, \phi,\psi)$ as

\begin{footnotesize}
\begin{equation} \label{eq:Euler angles}
\left( \begin{array}{c c c}
    \cos(\phi)\cos(\psi) - \cos(\theta)\sin(\psi)\sin(\phi)& -\sin(\phi)\cos(\psi)-\cos(\theta)\sin(\psi)\cos(\phi)& \sin(\theta)\sin(\psi)\\
    \cos(\phi)\sin(\psi) + \cos(\theta)\cos(\psi)\sin(\phi)& -\sin(\phi)\sin(\psi)+\cos(\theta)\cos(\psi)\cos(\phi)& -\sin(\theta)\cos(\psi)\\
   \sin(\theta)\sin(\phi)& \sin(\theta)\cos(\phi)& \cos(\theta)
    \end{array}\right)
\end{equation}
    \end{footnotesize}
and any element of $\mathfrak{so}(3)$ as
\begin{equation}
    \xi = \left(\begin{array}{c c c}
    0& -\boldsymbol\omega_3 & \boldsymbol\omega_2\\
    \boldsymbol\omega_3&0&-\boldsymbol\omega_1\\
    -\boldsymbol\omega_2&\boldsymbol\omega_1&0
    \end{array} \right)
\end{equation}

Now, consider the two points in $S^2\subset\mathbb{R}^3$, separated by a geodesic distance $q$:
\begin{equation}
\label{eq:initial points}
     \mathbf{x}_1 = \left(\begin{tabular}{c} 0\\ 0\\-1\end{tabular}\right), \quad \mathbf{x}_2 = \left(\begin{array}{c} 0\\\sin(q)\\-\cos(q) \end{array}\right)
\end{equation}
It is straightforward to see that any configuration can be obtained by placing the two particles at an appropriate distance in the form above and then rotating them. Hence, the tangent bundle can be rewritten in terms of the angles $\theta, \psi, \phi$, the coordinates on the Lie algebra $\boldsymbol\omega_1,\boldsymbol\omega_2,\boldsymbol\omega_3$, the distance $q$ and its rate of change $\dot{q}$.

Due to the nature of our phase space, we can assume that  that $q\in I:=(0,\pi)$. Left trivialization of $T\mathsf{SO}(3) = \mathsf{SO}(3)\times\mathfrak{so}(3)$ allows us to  write the parametrization
\begin{equation}
\label{eq: spherical coords two particleS}
    \begin{split}
       & TI\times T\mathsf{SO}(3)\to T\mathcal{Q}\\
        &(q,\dot{q}, \theta,\phi, \psi,\boldsymbol\omega_1,\boldsymbol\omega_2,\boldsymbol\omega_3)\mapsto(g\cdot \mathbf{x}_1, g\cdot \mathbf{x}_2, g\xi\cdot\mathbf{x}_1, g\xi\cdot\mathbf{x}_2 + g\mathbf{x}_2'\dot{q}),
    \end{split}
\end{equation}
where $g$ is given in \eqref{eq:Euler angles}. This gives $\xi$ the physical meaning of  angular velocity in the body frame \cite{marsden2013introduction}.
 
Since the problem is not determined by a Lagrangian, to transfer to a set of conjugate variables we use a slightly modified method from the usual one (compare with the Lagrangian approach in \cite{borisov2018reduction}). 

Let $T$ denote  the kinetic energy of the system (as rewritten in our reduced coordinates): $T =  \frac{\mu_1}{2}
    (\boldsymbol\omega_1^2 + \boldsymbol\omega_2^2) + \frac{\mu_2}{2}\left((\boldsymbol\omega_1 + \dot{q})^2 + (\sin(q)\boldsymbol\omega_3 + \cos(q)\boldsymbol\omega_2)^2\right)$ and introduce  $m_i = \frac{\partial T}{\partial\boldsymbol\omega_i}$ in place of the $\boldsymbol\omega_i$.

   Moving from $TI$ to $T^{\ast}I$ can be accomplished by taking $p: = \frac{\partial T}{\partial\dot{q}}$.
    The relations between the variables are as follows:
\begin{equation}
\label{eq: omegas through ms}
 \left\{   \begin{array}{lcl}
    \boldsymbol\omega_1 &=&  \frac{1}{\mu_1}(m_1-p),\\[4pt]
   \boldsymbol \omega_2 &= & \frac1{\mu_1}(m_2-m_3\cot(q))\\[4pt]
    \boldsymbol\omega_3 &= &\frac1{\mu_1}(\cot(q)(m_3\cot(q) -m_2)) + \frac1{\mu_2}(m_3\csc^2(q))\\[4pt]
    \dot{q}  &= & \frac1{\mu_1\mu_2}\left(p (\mu_1+\mu_2)-\mu_2 m_1\right)
    \end{array}\right.
\end{equation}

The momentum map $\Phi$ is, in the new set of variables,
\begin{equation}
    \Phi(g,m_1,m_2,m_3,q,p) \ = \ g\cdot
    \begin{pmatrix}
    m_1\\ 
    m_2 - B e_2 \sin (q)\\ 
    m_3+B e_1 + B e_2 \cos (q)
    \end{pmatrix}.
\end{equation}

Since the symplectic form is not standard, we need to rewrite it according to the general rule: if the change of coordinates is given by a Jacobian matrix $J$, the new symplectic structure will be $\tilde{\omega} = (J)^T\omega J$. Keeping our choice of signs in line with \cite{littlejohn1979guiding}, we write the Poisson structure as $\tilde{\sigma} = -\left((J)^T\omega J\right)^{-1}$, and  the Hamiltonian equations are consequently given by $\tilde{\sigma}\cdot \mathrm{grad} \ \mathcal{H}$, with $\mathrm{grad} \ \mathcal{H}$  rewritten in the new set of variables.

After performing the calculations, it can be observed that the last five equations form an independent subsystem:

\begin{equation}
\label{redalt}
\left\{\begin{array}{lcl}
\dot{m}_1&=& -\frac{1}{\mu_1\mu_2}\left(\mu_2 (m_2-m_3 \cot q) (Be_1+m_2 \cot q+m_3)+Be_2 \mu_1 m_3 \csc q-\mu_1 m_2 m_3 \csc ^2q\right)\\[6pt]
\dot{m}_2 &=& \frac{1}{\mu_1\mu_2}\left(\mu_2 (m_1-p) (Be_1+m_3)+Be_2 \mu_1 p \cos (q)+\right.\\
&&\qquad\qquad\left.+\mu_2 m_1 \cot q (m_2-m_3 \cot q)-\mu_1 m_1 m_3 \csc ^2q\right)\\[6pt]
\dot{m}_3 &=&\frac{1}{\mu_1\mu_2}\left(\mu_1(Be_2 p \sin q)+\mu_2(m_2 p-m_1 m_3 \cot q)\right)\\[6pt]
\dot{q}  &=& \frac{1}{\mu_1\mu_2}\left(p (\mu_1+\mu_2)-\mu_2 m_1\right)\\[6pt]
\dot{p} &=&-\frac{1}{\mu_1\mu_2}\left(m_3 \csc q (Be_2 \mu_1+\csc q (\mu_2 m_2-m_3 (\mu_1+\mu_2) \cot q))+\mu_1 \mu_2 V'(q)\right). 
\end{array} \right.
\end{equation}

The quintuple of coordinates $(m_1, \ m_2, \ m_3, \ q, \ p)$ describes the system reduced with respect to the $\mathsf{SO}(3)$-action. From here on, we  restrict our attention to these reduced equations.

The Hamiltonian is rotationally invariant, so its reduced form is obtained just by substitution of the new set of variables:   
\begin{small}

\begin{equation}
\label{hamilt red}
\begin{split}
\mathcal{H} \ = \ & \frac{1}{2\mu_1\mu_2 }\biggl(
\mu_2  \left((m_1-p)^2+m_2^2\right)+m_3 \left(-2 \mu_2  m_2 \cot (q)+\mu_1  m_3 \csc^2q+\mu_2  m_3 \cot^2q\right)\biggr)+{}\\&\qquad{}+\frac{ p^2}{2 \mu_2 } +  V(q).
\end{split}
\end{equation}\end{small}
The  non-zero Poisson brackets in the reduced variables are given by 
    \begin{align}
    \label{Poiss}
        &\{m_1,m_2\} = -m_3-B (e_1+ e_2 \cos (q)),\ &  &\{m_2,m_3\} = -m_1 ,\nonumber\\ &\{m_1,m_3\} = m_2-B e_2 \sin (q), \ & &\{m_2,p\} = B e_2 \cos (q) , \nonumber\\ &\{m_3,p\} = B e_2 \sin(q),\ &  &\{q,p\} = 1.
    \end{align}
It can easily be seen that the Poisson structure \eqref{Poiss} is generically of rank 4, and  has Casimir function given by the square of the momentum map:
\begin{equation}
\label{eq: Casimir reduced}
    \mathcal{C} = m_1^2 + (m_2- Be_2 \sin q)^2 +(m_3+B (e_1+e_2\cos q))^2 .
\end{equation}

\begin{rmk}
The limiting case of $B = 0$ takes these expressions to the reduced equations of motion for the two body  problem on a sphere from \cite{borisov2018reduction,garcia2020attracting} (for details, see Appendix).\end{rmk}
\begin{rmk}
\label{st:reverse}
 Note that the simultaneous change in the signs of $B, m_1,m_2,m_3,p$ is a time reversing symmetry, for it takes the Poisson structure to its opposite, while  leaving $\mathcal{H}$ invariant.
\end{rmk}

\subsection{Relative equilibria}
\label{sec:Relative equilibria}

Our primary goal is describing and locating  relative equilibria of the system. The condition for that is the right hand side of the system \eqref{redalt} equals  0.

Solving  \eqref{redalt} gives $p = m_1=0$ from the second and  fourth equations. Assuming otherwise leads to a linear relation between $m_2$ and $m_3$, which ultimately yields $V'(q) = 0$, contradicting our initial assumptions about the potential. 

Substituting zero values of $m_1$ and $p$ into the equations, we obtain a system for $m_2$ and $m_3$, consisting of the first and the fifth equation from \eqref{redalt}. 

Solving the first equation for $m_2$ yields

\begin{equation}
\label{sols m2}
m_2 = \frac1{2 \mu_2}\tan (q) \left(m_3(\mu_1\csc^2q+\mu_2 \cot^2q-\mu_2) -Be_1 \mu_2 \pm\sqrt{A}\right)
\end{equation}
where 
\begin{equation}
\begin{split}
\label{eq: A}
    A \  =\  4 \mu_2 m_3 &\cot (q) \csc (q) (\mu_2 \cos (q) (Be_1+m_3)-Be_2 \mu_1)+{}\\&{}+\Bigl(-\mu_2 (Be_1+m_3)+\mu_1 m_3 \csc^2q+\mu_2 m_3 \cot^2q\Bigr)^2.
    \end{split}
\end{equation}

Firstly, we note that the expression for $m_2$ is indeterminate when  $q=  \frac{\pi}{2}$, separating this into a special case.

Secondly, due to a square root being present in the expressions, the right hand side of \eqref{eq: A} is only defined for some values of $m_3$. The inequality
\begin{equation}
\label{permitted}
    \begin{split}
       A \ge 0
    \end{split}
\end{equation}
describes the permitted values of the variable.

A little rearranging  of the polynomial \eqref{eq: A} shows that the coefficient of $m_3^2$ is  
$$\csc ^4(q) \left(\mu_1^2+\mu_2^2+2 \mu_1\mu_2 \cos (2 q)\right).$$ 
Note that since this expression  is always positive, all the `bad' values of $m_3$ in \eqref{permitted} always lie between the two solutions of the equation $A = 0$.

Substituting  \eqref{sols m2} into the fifth equation of \eqref{redalt}, we obtain, for $m_3$:
\begin{small}
\begin{equation}
\label{eq:that one with breaks}
    m_3 \csc (q) \sec (q) \left(B e_1\mu_2-2 Be_2\mu_1 \cos (q)-2 \mu_1  m_3+m_3 (\mu_1+\mu_2) \csc^2q\pm\sqrt{A}\right) - \ 2 \mu_1 \mu_2 V'(q)=0.
\end{equation}
\end{small}
In order to prove the existence of roots for \eqref{eq:that one with breaks}, we employ the following observation: the value of $A$ at $m_3 = 0$ is $B^2 e_1^2 \mu_2^2$, whence $A$ is always positive at $m_3 = 0$. Thus, the interval in $m_3$ where the function on the left hand side of \eqref{eq:that one with breaks} is not defined lies entirely to the left or to the right of $m_3=0$. 

Consider the  two equations in \eqref{eq:that one with breaks}. With appropriate arrangements, they both square to a fourth degree polynomial in $m_3$:
\begin{small}
\begin{equation}
    \label{eq:the  one in m3}
    \begin{split}
    &-4 \mu_1 m_3^4 \csc ^4(q)  + 4 B  m_3^3 \csc ^4(q) (\text{$e_1$} \mu_2-\text{$e_2$} \mu_1 \cos (2 q) \sec (q))+ \\&+2  m_3^2 \csc ^3(q) \sec (q) \left(2 B^2e_2 \sin (q) (\text{$e_2$} \mu_1 \cos (q)-\text{$e_1$} \mu_2)-2 \mu_2 V'(q) (\mu_2+\mu_1 \cos (2 q))\right)+ \\&+  4 B \mu_2 m_3 \csc (q) V'(q) (2e_2 \mu_1-\text{$e_1$} \mu_2 \sec (q)) + 4 \mu_1 \mu_2^2 V'(q)^2 = 0
    \end{split}
\end{equation}
\end{small}
At this point we employ a classical theorem:

\paragraph{Descarte's rule of signs \cite{descartes1886geometrie}:}{\it
The number of positive roots of a polynomial with real coefficients is equal to or less by an even number than the number of changes of sign of the coefficients of the polynomial in question, when written in the order of descending degree of the variable.
}

\bigskip

By looking at   \eqref{eq:the  one in m3}, it can be  easily observed that since $\mu_1>0, \mu_2>0$, the highest coefficient is always negative, and the lowest one is positive. Since the degree of the polynomial is 4, Descarte's rule of signs, as applied to the polynomial of $m_3$ and then $-m_3$, gives that the first equation in \eqref{eq:the  one in m3} must have at least one positive and one negative root. But then at least one of them does not lie in the interval between the roots of \eqref{eq: A}. Hence, for \eqref{eq:that one with breaks}, at least one solution always exists.

Now, we fill in the gap  by taking $q = \frac{\pi}{2}$ where we can give a more precise statement depending on the value of $V'(\pi/2)$. Taking $m_1 = 0, \ p = 0, \  q = \frac{\pi}{2}$ in the system  \eqref{redalt} gives us only two nontrivial equations:
\begin{equation}
\label{q = pi/2}
\begin{array}{lcl}
\mu_2 m_2 (Be_1+m_3)+Be_2 \mu_1 m_3-\mu_1 m_2 m_3 &=& 0,\\
m_3 (Be_2 \mu_1+\mu_2 m_2)+\mu_1 \mu_2 V'\left(\frac{\pi }{2}\right) &=& 0
\end{array}
\end{equation}

By analytically solving \eqref{q = pi/2}, one can check that it has real solutions if and only if 
\begin{equation}
\label{cond for q  = pi/2}
   B^4e_1^2e_2^2+2 B^2e_1e_2 (\mu_1+\mu_2) V'\left(\frac{\pi }{2}\right)+(\mu_1-\mu_2)^2 V'\left(\frac{\pi }{2}\right)^2\ge0, 
\end{equation}
with two solutions if the left hand side of \eqref{cond for q  = pi/2} is strictly greater than 0, one if it is equal to 0 and none when it is less than zero. 
\begin{rmk}
\label{rem: limiting case solutions}
 From the explicit form of the solutions of \eqref{q = pi/2} (we omit the computations here), it can be easily seen that the limiting case $B= 0$ agrees with the results for the two body problem on a sphere from \cite{borisov2018reduction} and \cite{garcia2020attracting}: no relative equilibria exist when $q = \frac{\pi}{2}$ unless $\mu_1 = \mu_2$. 
\end{rmk}
Thus, we have demonstrated the

\begin{theorem}
\label{st: Existence of eq gen case} 
\begin{enumerate}
\item For each $q\in(0,\pi)\setminus\{\pi/2\}$, and for all non-zero values of $B,\ e_1,\ e_2, \ \mu_1,\ \mu_2$ and any smooth function $V(q)$ there exists at least one relative equilibrium;
\item for $q=\pi/2$, there are precisely 0, 1 or 2 relative equilibria accordingly as the discriminant in \eqref{cond for q  = pi/2} is negative, zero or positive.
\end{enumerate}
\end{theorem}

\section{Identical particles}

The most natural case to investigate closely is the case of two identical particles, that is the one with  $e_1 = e_2 = 1, \  \mu_1= \mu_2 = 1$. Without loss of generality we may assume $B>0$, as  $B<0$ can be reduced to this via the time-reversing symmetry described in Remark \ref{st:reverse}.  In this section we describe both the existence and stability of the relative equilibria for identical particles (in Section\,\ref{sec:opposite charges} we consider the setting where the particles have the same mass but opposite charges).

Our choice of the potential is $V(q) = \cot(q)$. This gives $V'(q) = -\csc^2(q)<0$, so describes a repelling force in accordance with the physics.

The reduced equations of motion \eqref{redalt} then read:
\begin{equation}
    \left\{\begin{array}{lcl}
    \label{two id part equat}
    \dot{m}_1 &=& -(m_2-m_3 \cot (q)) (B+m_2 \cot (q)+m_3)-m_3\csc(q)(B -m_2  \csc (q)),\\[6pt]
     \dot{m}_2 &=&m_1 (B+2 m_3)-p (2B\sin^2\left(\frac{q}{2}\right)+m_3)+m_1 m_2 \cot (q)-2 m_1 m_3 \csc^2q,\\[6pt]
      \dot{m}_3 &=&p (B \sin (q)+m_2)-m_1 m_3 \cot (q),\\[4pt]
       \dot{q} &=&  2 p-m_1, \\[4pt]
        \dot{p} &=& \csc (q) \left(\csc (q) \left(-m_2 m_3+2 m_3^2 \cot (q)+1\right)-B m_3\right).\\
      \end{array}\right.
\end{equation}

\subsection{Classification of relative equilibria}

We proceed  to find stationary points of the system \eqref{two id part equat}, which correspond to relative equilibria of the original system.  The conditions $\dot{m}_2=\dot{m}_3=\dot{q}=0$ imply $p=m_1=0$ (as pointed out above for the more general case).  If we then solve $\dot{m}_1=0$ for $m_2$ and substitute the two solutions of this into $\dot{p}=0$ we obtain two quadratic equations (since $A$ from \eqref{eq: A} becomes a square)  for $m_3$.  The four solutions can be split into 2 pairs leading to the following solutions (the analytic expressions were found using \textsc{Mathematica}).
\begin{align}
  \mathrm{Type\ I}:\ \quad
 & \left\{\begin{array}{lcl}
 \label{eq:idensolsa}
    m_1 &=& 0,\\ p &= &0,\\ 
    m_2^{\pm} &= &\frac{2B\sin^3\left(\frac{q}{2}\right)\sin(q) \pm \cos^{\frac{3}{2}}\left(\frac{q}{2}\right)\cos(q)\sqrt{4\csc\left(\frac{q}{2}\right) + B^2\sec\left(\frac{q}{2}\right)\sin^2(q)\tan^2(q)}}{\sin\left(\frac{q}{2}\right) - \sin\left(\frac{3q}{2}\right)},\\[16pt]
    m_3^{\pm} &=& \frac{1}{2}\left(B\sin(q)\tan(q) \mp\sqrt{\cos\left(\frac{q}{2}\right)}\sqrt{4\csc\left(\frac{q}{2}\right) + B^2\sec\left(\frac{q}{2}\right)\sin^2(q)\tan^2(q)}\right);
    \end{array}\right. \\
      \mathrm{Type\ II}:\quad
   & \left\{\begin{array}{lcl}
    \label{eq:idensolsb}
    m_1 &= &0,\\     p &=& 0,\\ 
    m_2^{\pm} & = & -2 \sin ^4\left(\frac{q}{2}\right) \csc (q) 
    		\left(B\pm \sqrt{B^2-2 \csc ^2\left(\frac{q}{2}\right) \csc (q)}\right),\\[8pt] 
    m_3^{\pm} &= & \sin ^2\left(\frac{q}{2}\right) \left(B\pm\sqrt{B^2-2 \csc ^2\left(\frac{q}{2}\right) \csc (q)}\right).
    \end{array}\right.
\end{align}

\medskip

It is convenient to express the existence and stability of all the solutions in  \eqref{eq:idensolsa} and \eqref{eq:idensolsb} through the two  `parameters', $B$ and $q$. 

Observe that, except for $q=\pi/2$, the two solutions in \eqref{eq:idensolsa} exist for all  values of $B$ and $q$ --- we refer to these as relative equilibria of Type I. It turns out (see Lemma\,\ref{st:particles swap} below) that the two Type I solutions are related by exchanging the (identical) particles. 

\begin{figure}
\centering
{\includegraphics[scale=1.3]{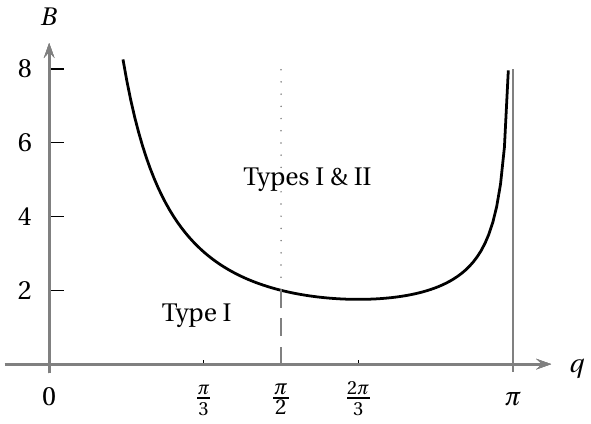}}
\caption{ The threshold  curve: only Type I solutions exist below the curve.}
\label{bif pic}
\end{figure}

For solutions \eqref{eq:idensolsb} to exist, the expression under the square root needs to be positive. It is clear that for each value of $q\in(0,\pi)$ when $B$ is sufficiently large, there is a solution.  We call these relative equilibria of Type II.
More precisely,  relative equilibria of Type II  exist when $B^2\geq2\csc^2(q/2)\csc(q)$.  Since we are assuming that $B>0$, the threshold value of $B$ is,
\begin{equation}
\label{b of q}
B = 2\sqrt{\frac{\csc(q)}{1- \cos(q)}}.
\end{equation}
We will refer to the graph of \eqref{b of q}, shown in Figure \ref{bif pic}, as "the threshold curve".

The minimal value that the function above assumes is $B  = \frac{4}{3}3^{1/4}$ (which is approximately $1.755$) at  $q= \frac{2\pi}{3}$. The value of \eqref{b of q} at $q = \frac{\pi}{2}$ is 2 (this fact will come useful later).

Note that for the values of $(q,B)$ on the threshold curve the two solutions from \eqref{eq:idensolsb} coincide. 

We summarize the above discussion in the following existence theorem. 

\begin{theorem} 
For the system described above, with two identical particles, we have the following relative equilibria for values of $B>0$: 
\begin{enumerate}
\item for every $q\in(0,\pi)\setminus\left\{\frac{\pi}{2}\right\}$ there is a unique relative equilibrium of Type I, up to particle exchange, and none exists for $q= \frac{\pi}{2}$;

\item 
\begin{enumerate}
    \item For  $0<B<\frac{4}{3}3^{1/4}$ there are no relative equilibria of Type II. 
    \item  Let $B=\frac{4}{3}3^{1/4}$. Then for $q= \frac{2\pi}{3}$ there is one relative equilibrium of Type II, while for $q\neq2\pi/3$ there are none. 
    \item Let $B>\frac{4}{3}3^{1/4}$. Then for $q\in(q_0,q_1)$  there are two distinct relative equilibria of Type II, for $q\in\{q_0,q_1\}$ there is just one while for $q\not\in [q_0,q_1]$ there are none, where $q_0,q_1$ denote the two solutions of\/ $2\sqrt{\frac{\csc(q)}{1- \cos(q)}}=B$ ($q_0\le q_1$).
\end{enumerate}
\end{enumerate}
\end{theorem}

\begin{rmk}
For an arbitrary choice of the potential $V(q)$ with $V'(q) \ne 0$ there are four solutions as well, with one pair existing for all values of $B$ and $q$, and the second pair for the values above some threshold curve.  
\end{rmk}

Before proceeding further with the analysis, we determine the effect of swapping the particles on the reduced space. 

\begin{lemma}
\label{st:particles swap}
The $\mathbb{Z}_2$-action of swapping the two identical particles is a symmetry of the system and induces a coordinate change on the reduced phase space which leaves $q$ invariant and multiplies $(m_1,m_2,m_3,p)^T$ by the matrix
 \begin{equation}
\label{particle change matrix reduced coordinate}
\begin{pmatrix}
 -1 & 0 & 0 & 0 \\
 0 & -\cos (q) & -\sin (q) & 0 \\
 0 & -\sin (q) & \cos (q) & 0 \\
 -1 & 0 & 0 & 1 \\
\end{pmatrix}.
\end{equation}.
\end{lemma}

\begin{proof}
Without loss of generality we can assume that the initial placement of our particles, and the one at which the exchange happens, is at the two points $\mathbf{x}_1$ and $\mathbf{x}_2$ from \eqref{eq:initial points}.

The matrix exchanging said points is given by
$$
   \mathsf{SO}(3) \ni A_0 := \begin{pmatrix}
    -1 & 0& 0\\
    0& -\cos(q)& -\sin(q)\\
    0& -\sin(q)& \cos(q)
    \end{pmatrix}.
$$

With this in mind,  the swapping can be written as $g\mathbf{x}_1\mapsto gA_0\mathbf{x}_1$, with a similar relation for the second particle. Therefore, under this $\mathbb{Z}_2$-action we have  for $\dot{g}$ and $\xi$
\begin{equation*}
\begin{split}
    g&\mapsto gA_0,\\
    \dot{g}&\mapsto \dot{g}A_0 + g\dot{A}_0 = g\xi A_0 + gA_0\xi_0 = gA_0(A_0^{-1}\xi A_0 + \xi_0),
    \end{split}
\end{equation*}
where $\xi_0$ is the tangent element of the Lie algebra $\mathfrak{so}(3)$ to the one-parameter subgroup $A_0$ with varying $q$.

Hence, 
$$    \xi\mapsto A_0^{-1}\xi A_0 + \xi_0$$
or, explicitly, 
$$
\begin{pmatrix}
0& -\boldsymbol\omega_3&\boldsymbol\omega_2\\
\boldsymbol\omega_3&0&-\boldsymbol\omega_1\\
-\boldsymbol\omega_2&\boldsymbol \omega_1&0
\end{pmatrix}\mapsto
\begin{pmatrix}
 0 & \text{$\boldsymbol\omega_2$} \sin (q)-\text{$\boldsymbol\omega_3$} \cos (q) & -\text{$\boldsymbol\omega_2$} \cos (q)-\text{$\boldsymbol\omega_3$} \sin (q) \\
 \text{$\boldsymbol\omega_3$} \cos (q)-\text{$\boldsymbol\omega_2$} \sin (q) & 0 & \dot{q}+\text{$\boldsymbol\omega_1$} \\
 \text{$\boldsymbol\omega_2$} \cos (q)+\text{$\boldsymbol\omega_3$} \sin (q) & -\dot{q}-\text{$\boldsymbol\omega_1$} & 0 \\
\end{pmatrix}
$$
The Jacobian matrix $\mathcal{J}_1$ of this change in the variables $(\boldsymbol\omega_1,\boldsymbol\omega_2,\boldsymbol\omega_3,\dot{q})$ is then
$$\begin{pmatrix}
 -1 & 0 & 0 & -1 \\
 0 & -\cos q & -\sin q & 0 \\
 0 & -\sin q & \cos q & 0 \\
 0 & 0 & 0 & 1 \\
\end{pmatrix},
$$
while the Jacobian $\mathcal{J}_2$ of the transfer from $(\boldsymbol\omega_1,\boldsymbol\omega_2,\boldsymbol\omega_3,\dot{q})$ to $(m_1,m_2,m_3,p)$ in the case of identical particles  is given by
$$
\begin{pmatrix}
 1 & 0 & 0 & -1 \\
 0 & 1 & -\cot (q) & 0 \\
 0 & -\cot (q) & \cot^2q+\csc^2q & 0 \\
 -1 & 0 & 0 & 2 \\
\end{pmatrix}.
$$
Finally, the final transformation matrix for the reduced conjugate coordinates is $\mathcal{J}_2^{-1}\mathcal{J}_1\mathcal{J}_2$ and is explicitly given by \eqref{particle change matrix reduced coordinate}, which is straightforwardly an involutory matrix.
\end{proof} 

We apply Lemma \ref{st:particles swap} to the Type I relative equilibria  
and multiply the vector $(0, m_2^+, m_3^+,0)$ by \eqref{particle change matrix reduced coordinate}, which maps it to $(0, m_2^-, m_3^-,0)$, and vice versa. Thus, the two solutions are the same up to changing the labelling of the particles, as claimed above.

\subsection{Variation of configurations for Type I relative equilibria}

Here we briefly comment on how the relative equilibria of the non-magnetic 2 body problem on the sphere with a repelling potential  \cite{garcia2020attracting} deform into the relative equilibria of Type I in the current problem, as mentioned in Remark \ref{rem: limiting case solutions}.

Let $\theta_1$ be the angle between the directed axis of rotation and the first particle  and $\theta_2$ the angle between that axis and the second particle.  For given $q$ we have $\theta_2=\theta_1+q$. 
 
\begin{figure}
\centering
 \subfigure[$B=0$]{
    \includegraphics[scale = 0.45]{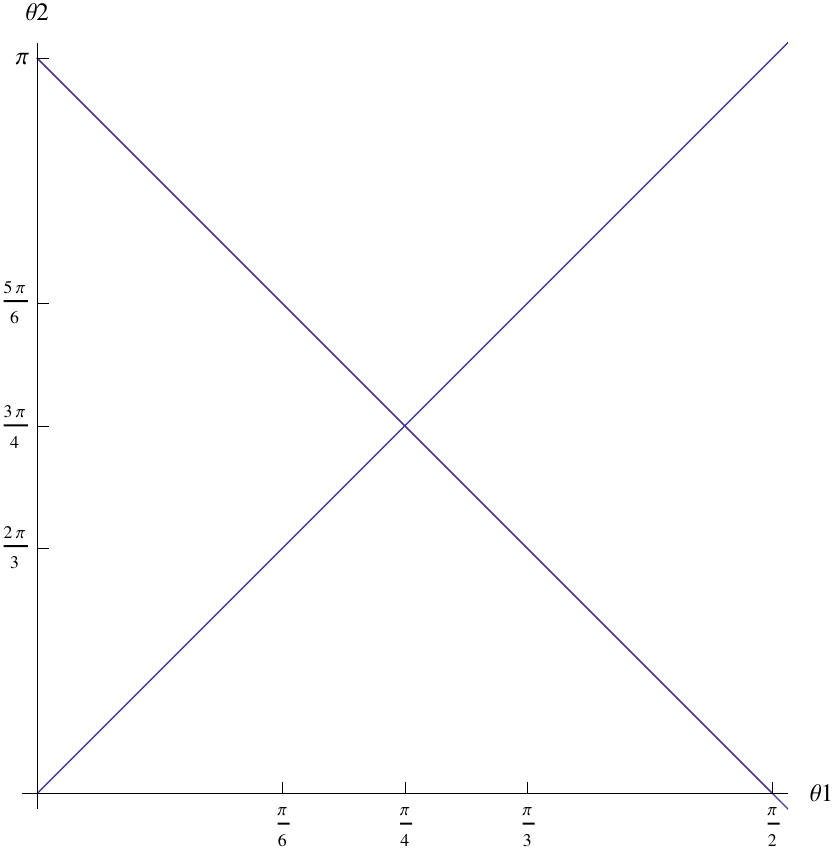}
    }
    \hfill
\subfigure[$B = 0.009$]{
\includegraphics[scale = 0.45]{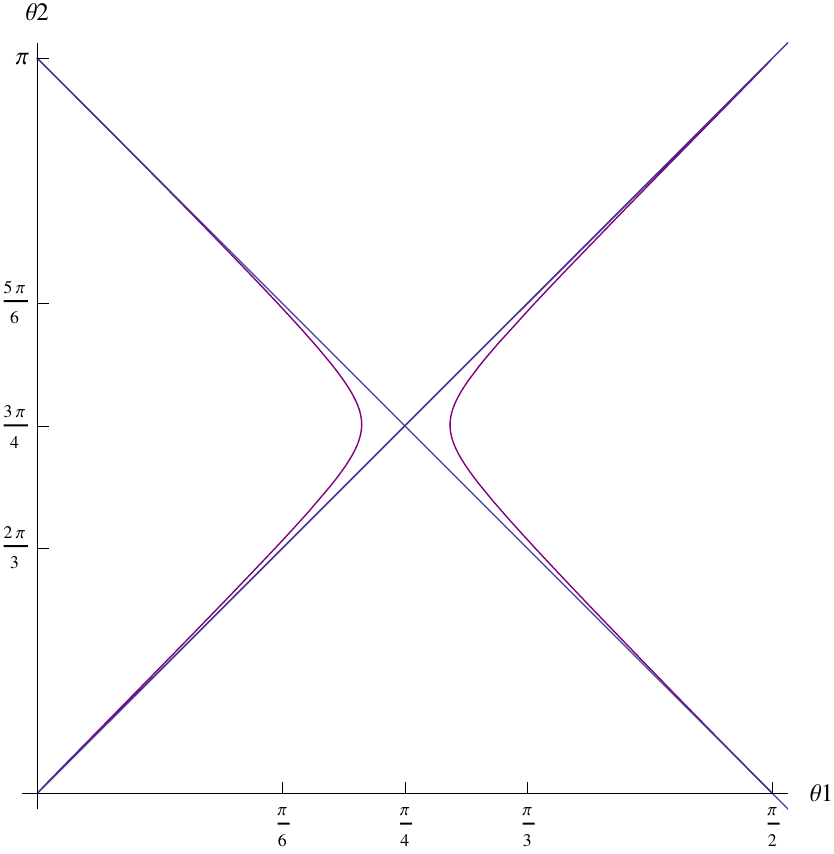}
}
\hfill
\subfigure[$B = 0.3$]{
\includegraphics[scale = 0.45]{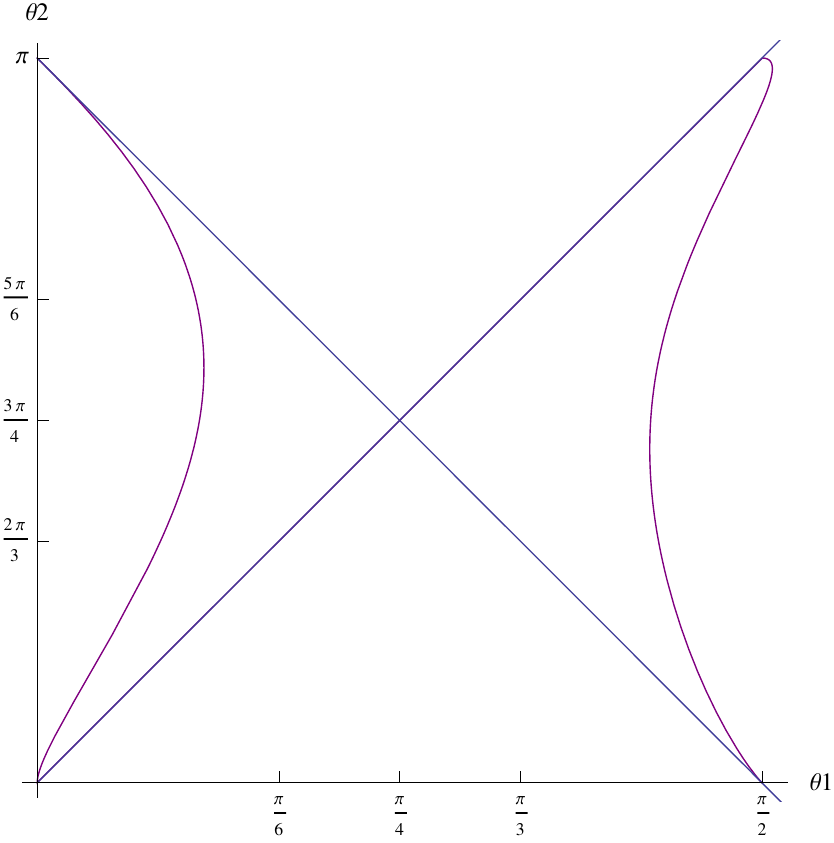}
}
\caption{Relative equilibria of Type I, plotted on $(\theta_1, \theta_2)$-plane for various values of $B$. The plot of relative equilibria for $B=0$ is provided in blue on the second and third picture for comparison.}
 \label{fig:Type I deformation}
\end{figure}

With $B=0$ there are two families of relative equilibria, the isosceles ones with $\theta_1+\theta_2=\pi$ and the right angled ones with $\theta_2-\theta_1=q=\pi/2$ \cite[Theorem 2.1]{garcia2020attracting}. These are the two lines shown in Figure\,\ref{fig:Type I deformation}(a). 
 
Now with $B>0$, for every $q \ \ne\frac{\pi}{2}$ there will be a single relative equilibrium of Type I (as mentioned previously, up to a particle exchange). Therefore, $\theta_1= \theta_1(q)$ and $\theta_2=\theta_1(q)+q$. This relation is plotted on $(\theta_1,\theta_2)$-plane  in Figure \ref{fig:Type I deformation}(b,c).

For negative values of $B$, the two branches of the curve will lie in the remaining upper and lower  quadrants formed by the lines in Figure \ref{fig:Type I deformation}(a).

\subsection{Reconstruction of motion}
Explicit formulae for relative equilibria in reduced coordinates allow us to reconstruct the motion of the particles (see Figure \ref{fig:acute and obtuse RE} for identical particles and Figure \ref{fig:RE opposite charges} for the case with opposite charges). By the nature of relative equilibria, said motion will be a rigid body type rotation of the particles around some fixed axis. 

 $\boldsymbol{\omega}_1$, \ $\boldsymbol{\omega}_2$, \ $\boldsymbol{\omega}_3$ from (\ref{eq: spherical coords two particleS}) are the components of  the body frame angular velocity vector of our system, which is parallel to the axis of rotation. Substituting the expressions for relative equilibria into (\ref{eq: omegas through ms}), assuming that one particle is at the point $(0,0,-1)$ and the other at $(0, \sin(q),-\cos(q))$ (namely, the body frame) gives us the desired picture. 

We explain the calculations in detail for relative equilibria of Type II: they can be repeated verbatim for Type I. Let us denote the angular velocity for $(m_2^{\pm}, m_3^{\pm})$ by $(\boldsymbol{\omega}_1^{\pm},\boldsymbol{\omega}_2^{\pm}, \boldsymbol{\omega}_3^{\pm}) $. It is straightforward that  $\boldsymbol{\omega}_1^{\pm} = 0$. 
The two other coordinates two are given by  

\begin{equation*}
\begin{cases}
 \boldsymbol{\omega}_2^+ =-\frac{1}{2} \tan \left(\frac{q}{2}\right) \left(B+\sqrt{B^2-2 \csc ^2\left(\frac{q}{2}\right) \csc (q)}\right),\\
     \boldsymbol{\omega}_3^+ = \frac{1}{2} \left(B+\sqrt{B^2-2 \csc ^2\left(\frac{q}{2}\right) \csc (q)}\right);
    \end{cases}
    \begin{cases}
    \boldsymbol{\omega}_2^- =-\frac{1}{2} \tan \left(\frac{q}{2}\right) \left(B-\sqrt{B^2-2 \csc ^2\left(\frac{q}{2}\right) \csc (q)}\right),\\
     \boldsymbol{\omega}_3^- = \frac{1}{2} \left(B-\sqrt{B^2-2 \csc ^2\left(\frac{q}{2}\right) \csc (q)}\right).
    \end{cases}
    \end{equation*}

 It is easy to check that the two vectors $(\boldsymbol{\omega}_2^{+}, \boldsymbol{\omega}_3^{+})$ and $(\boldsymbol{\omega}_2^{-}, \boldsymbol{\omega}_3^{-})$ are parallel. One can easily  deduce that  the cosines of the angles  between the axis of rotation and the two vectors $(0,-1)$ and $(\sin(q),-\cos(q))$ are equal (both of them are $\cos(q/2)$).  
 Hence, we can conclude that relative equilibria of Type II are isosceles configurations.  (This also follows from the fact that the reduced equilibria are fixed by the particle exchange symmetry.) 
 
Identical manipulations with the formulae for the relative equilibrium of Type I give that the difference between $\cos(\theta_1)$ and $\cos(\pi- \theta_2)$ (the minimal angle between the axis of rotation and the coordinate vector of the second body) will be equal to ${B (\sec (q)+1)}/({\sqrt{B^2 \sec ^2(q)+2 \csc ^3(q))}}$, a function that is zero if and only if $B=0$. Therefore,  the two minimal angles between coordinate vectors and the axis of rotation are not equal unless the magnetic field is absent.

\subsection{Energy-Casimir map}

Having the solutions of \eqref{two id part equat} of Types I and II, we want to approach the problem from a more physical angle: that of the energy-Casimir (or energy-momentum) map.

The first natural question to ask is that of the form of its image: namely, what does the set of values of $(\mathcal{C}, \mathcal{H})$ look like? Since $\mathcal{C}\geq0$, it lies  entirely in the right half-plane, including the boundary $\mathcal{C}=0$.  

To determine which values $\mathcal{H}$ can assume with a fixed  $\mathcal{C} = C_0$, we assign  specific values to our variables: $m_1= \sqrt{C_0},\ m_2 = B\sin(q),\  m_3 = -B(1 + \cos(q)),\  p=0$. At all of these points $\mathcal{C}$ is indeed equal to $C_0$, and $\mathcal{H}$ is a function of $q$, reduced to 
\begin{equation}
\label{eq Ham all values}
    \mathcal{H}(q) = \frac12 C_0 + \cot(q) +B^2\cot^2\left(\frac{q}{2}\right),
\end{equation}
This is a simple monotonic decreasing function of $q$, with limits $+\infty$ when $q\to 0$ and  $-\infty$ when $q\to \pi$.
Therefore, \emph{the image of the energy-momentum map is the entire right half-plane in} $(\mathcal{C},  \mathcal{H})$.

\paragraph{Zero level set of the Casimir}

Because of the magnetic term in the momentum map or Casimir, it could be particularly interesting to ask about the zero level set.  However, it turns out not to be so interesting!  

The zero level set of the Casimir is given by $m_1 = 0,\ m_2 = B\sin( q), m_3 = -B(1 + \cos (q)$. 
After substituting these values (along with $p=0)$ into the system \eqref{two id part equat}, we obtain one equation in $q$ for equilibrium points:
\[
\csc(q) + 2B^2\cot^2\left(\frac{q}{2}\right) = 0,
\]
which clearly has no solutions when $q\in\left(0,\pi\right)$. Therefore, as Figure \ref{fig:Ham Cas map image} suggests, there are no equilibria on the zero level set of the Casimir.

The reduced system is described in terms of $p$ and $q$ only, and thus is integrable with just the Hamiltonian, which has the form 
\begin{equation}
\label{eq: Hamiltonian when C=0}
   \mathcal{H}(q, p) =  \frac{3}{2}p^2 + B^2\cot^2\left(\frac{q}{2}\right) + \cot(q).
\end{equation}
The level sets of this function are non-compact, and consequently all motion is unbounded

\subsubsection{Energy-Casimir bifurcation diagram} \label{sec:Bifurcation diagram1}
\begin{figure}
\centering
\subfigure[A larger scale view]{\includegraphics[scale =0.5]{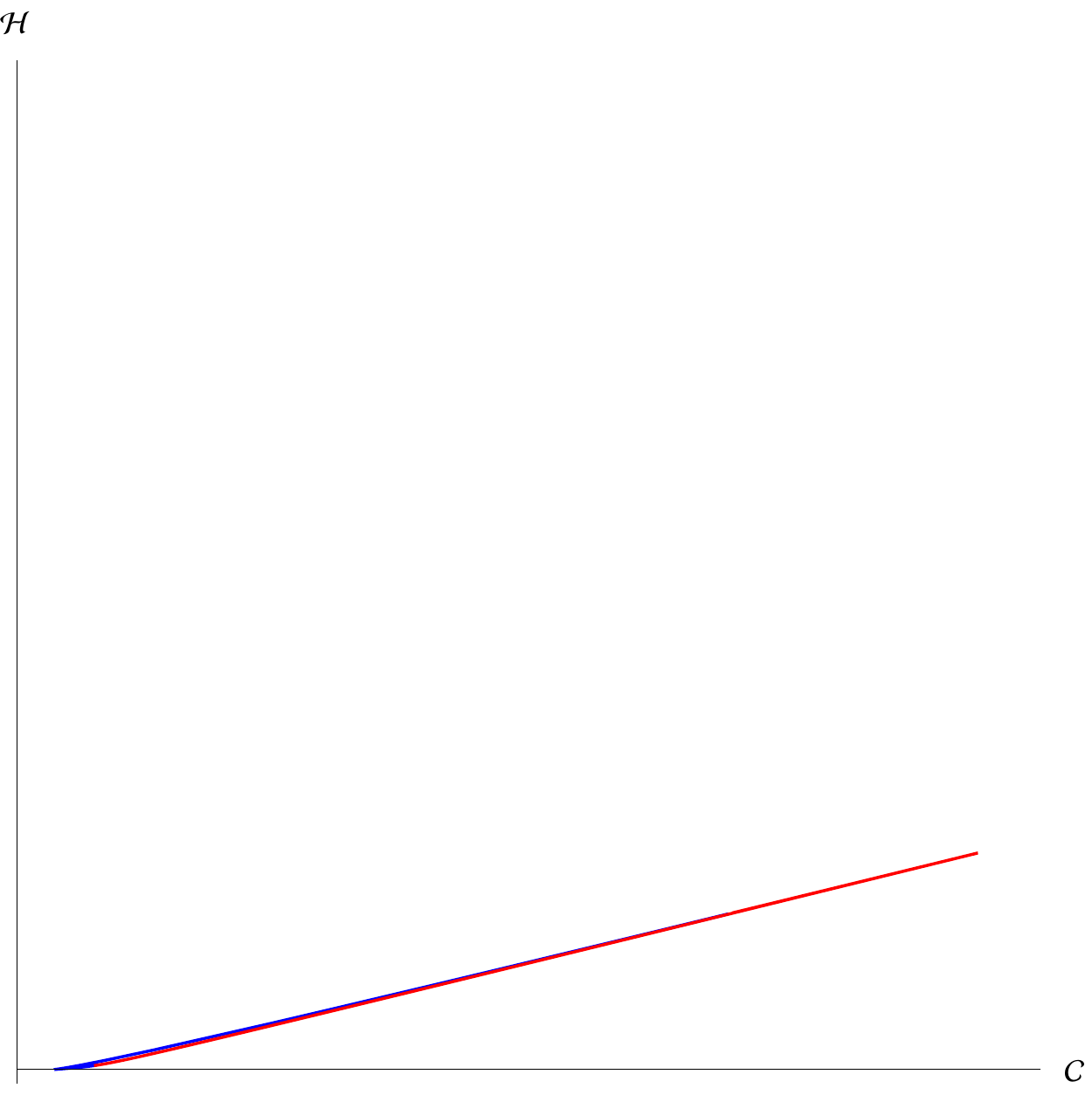}}
\hfil
\subfigure[A close-up; the continuations of red and blue lines form a closed loop, as in (a)]{\includegraphics[scale =0.5]{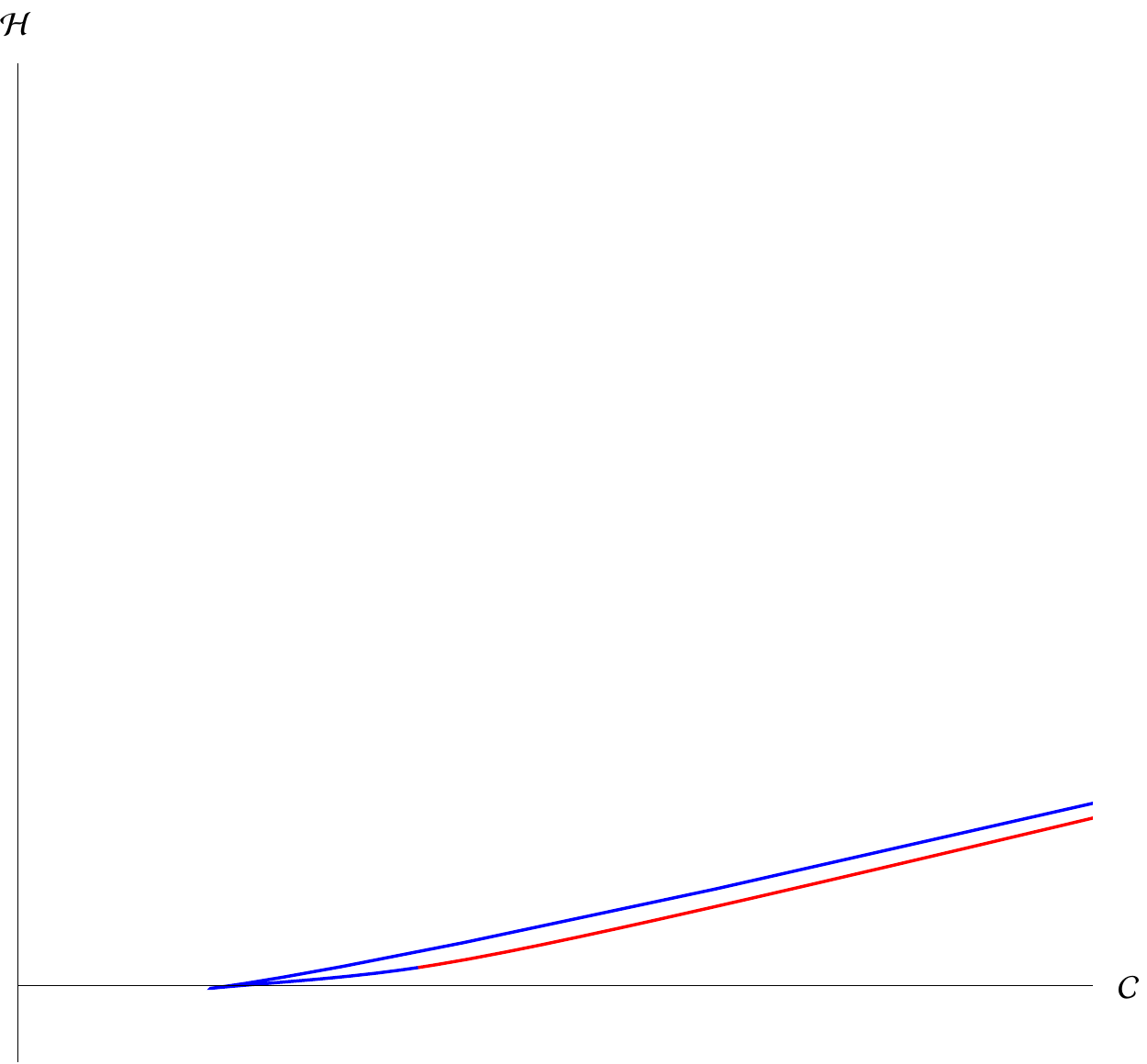}}
     \caption{Energy-momentum bifurcation diagram with $B = 2.5$; the red and blue curve corresponds to Type II relative equilibria, the others to Type I. See also Fig.\,\ref{fig:schematic energy-momentum}}
     \label{fig:Ham Cas map image}
\end{figure}

Figure \ref{fig:Ham Cas map image}  illustrates the set of singular values of the energy-Casimir map for a fixed value of $B$, which form the `bifurcation diagram' (see also Figure\,\ref{fig:schematic energy-momentum}). It shows a relatively large-scale view and a close up of a neighbourhood of the origin.  Singular points of the energy-Casimir map are relative equilibria so the curves shown are the energy-Casimir values on the set of relative equilibria.  

Of the different curves shown, the upper most one consists of obtuse Type I relative equilibria.  Along this curve, as $\mathcal{C}\to0$ so $q\to\pi$ and $\mathcal{H}\to-\infty$ (we have seen above that there are no equilibria for $\mathcal{C}=0$), and as $\mathcal{C}\to\infty$ so $q\to\frac\pi2^+$ and $\mathcal{H}\to\infty$ .

The intermediate curve, with a cusp, represents acute Type I relative equilibria. When $q\to 0, \ \mathcal{H}\to +\infty$, making up the lower branch of this curve; when $q\to\frac{\pi}{2}^{-}, \ \mathcal{H}\to +\infty$, forming its upper branch.

The lowest curve shows the image of the Type II relative equilibria. What is not shown at this scale is that the two branches of the Type II relative equilibria meet again in a second cusp point.  If this diagram is shown at a scale which shows the entirety of the Type II relative equilibria, the two branches would be indistinguishable; for this reason we illustrate it with a schematic diagram in Figure\,\ref{fig:schematic energy-momentum}(b).  The blue and red parts of this curve correspond to $(m_2^-, m_3^-)$ and $(m_2^+, m_3^+)$ from \eqref{eq:idensolsb}, respectively.

\def\cC{\mathcal{C}}
\def\cH{\mathcal{H}}
From the point of view of the reduced systems, with the Casimir $\mathcal{C}$ as the natural parameter, the behaviour is as follows.  For small values of $\mathcal{C}$, the reduced space has just one (relative) equilibrium: an obtuse Type I configuration (which persists for all values of $\mathcal{C}$);   as $\cC$ increases (assuming $B>\frac43 3^{1/4}$), two Type II relative equilibria appear from a saddle-node bifurcation and then, increasing $\mathcal{C}$ further there appears a pair of acute Type I relative equilibria, also in a saddle-node bifurcation. For a still larger value of $\mathcal{C}$ the two Type II relative equilibria merge in yet another saddle-node bifurcation, and no longer exist for large values of $\mathcal{C}$.  The precise values of $\mathcal{C}$ (and $q$) at which these three saddle-node bifurcations appear depends on the strength of the magnetic field.

We return to these bifurcation curves when looking at stability below.

\subsection{Stability of Type I relative equilibria} \label{sec:Type I stability}

A lengthy calculation shows that the characteristic polynomial of the linearized matrix of the system \eqref{two id part equat} is  of the form $x(-x^4 + ax^2 + b)$ (as it would be for every 4 dimensional Hamiltonian system: the factor of $x$ is due to the Casimir). Note that the coefficients $a$ and $b$ here are functions of $B$ and $q$.

Substituting $y = x^2$, we obtain a quadratic equation: $-y^2 + ay + b = 0$. Thus, for linear stability, both solutions of this equation must be negative. That is,  three conditions need to be fulfilled:
\begin{itemize}
    \item the top of the parabola must be to the left of $y = 0$ (i.e., $a<0$),
    \item the value of $-y^2 + ay + b$ at the top must be greater than 0 (i.e., $a^2+4b>0)$,
    \item the value of $-y^2 + ay + b $ at $y = 0$ must be less than 0 (i.e., $b<0$). 
\end{itemize}

In this case, the conditions for stability can be written analytically.

\begin{figure}
\centering
\includegraphics[scale = 0.8]{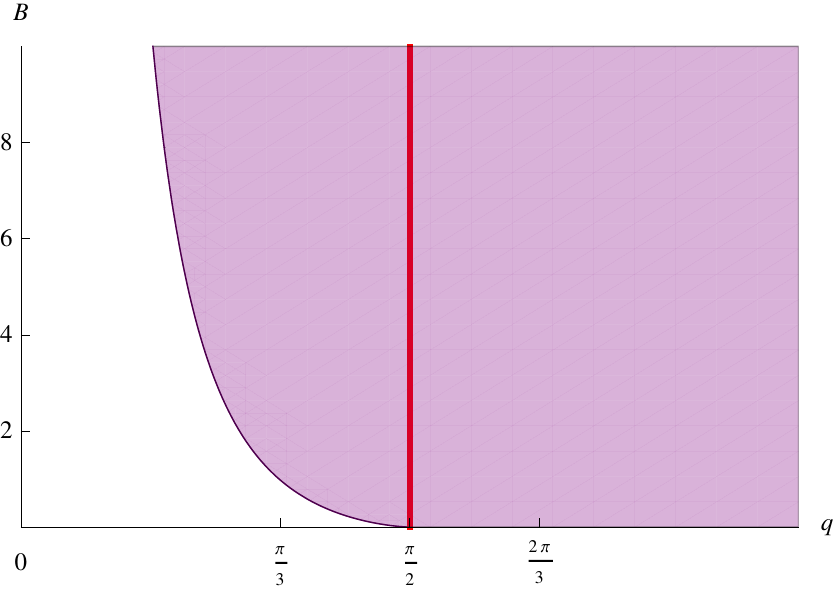}
\caption{Stability for the relative equilibria of Type I: the coloured region is where the relative equilibria are linearly stable}
\label{fig:Type I stability}
\end{figure}

Since the two relative equilibria of Type I are related by particle exchange (and the particles are identical) we only need perform calculations for one of the explicit solutions. By doing so, we  obtain Figure \ref{fig:Type I stability}. 

Relative equilibria are linearly stable in the region coloured lilac and linearly unstable in the white coloured one. The thick red line is where $q= \frac{\pi}{2}$ for which there is no relative equilibrium.  
 
In Figure \ref{fig:Type I stability}, the transition curve is obtained from the third condition for stability above ($b=0$) and is explicitly given by the expression   
$B = \sqrt{\frac{\cos^3(q)(2 + \cos(q))}{2\sin^3\left(q\right)\sin^2\left(\frac{q}{2}\right)}}$. Note that the graph meets the horizontal axis $B=0$ at $q = \frac{\pi}{2}$.

We have demonstrated

\begin{proposition}
For the system with identical particles, the linear stability of the relative equilibria of Type I  (solutions of \eqref{eq:idensolsa}), is as follows:
\begin{itemize}
   \item when $(q,B)$ is to the left of the graph  $B = \sqrt{\frac{\cos^3(q)(2 + \cos(q))}{2\sin^3\left(q\right)\sin^2\left(\frac{q}{2}\right)}}$, the relative equilibrium is linearly unstable;
     \item when $(q,B)$ is to the right of the graph  $B = \sqrt{\frac{\cos^3(q)(2 + \cos(q))}{2\sin^3\left(q\right)\sin^2\left(\frac{q}{2}\right)}}, q\ne \frac{\pi}{2}$, the  relative equilibrium is linearly stable.
\end{itemize}
See Figure\,\ref{fig:Type I stability}.
\end{proposition}

\begin{rmk}Almost all of the linearly stable Type I relative equilibria will probably be nonlinearly (Lyapunov) stable, by KAM theory; however there are non-degeneracy conditions to check in the higher order terms of the Hamiltonian near each relative equilibrium. For the non-magnetic 2-body problem these conditions are checked numerically for many of the relative equilibria \cite[Sec.\ 4.2.2]{borisov2018reduction}.   We do not pursue this here.
\end{rmk}

\subsection{Stability of Type II relative equilibria} \label{sec:Type II stability}

Here, we employ the same method as the one from the previous section. However, it has not been possible to obtain analytic results for  general values of $q$ and $B$ for these relative equilibria.

\subsubsection{Points on the threshold curve}

First, we investigate the stability on the  line $B = 2\sqrt{\frac{\csc(q)}{1- \cos(q)}}$, where one can obtain analytic results. Here, as previously mentioned, the two relative equilibria of Type II, namely $(m_2^\pm,m_3^\pm)$, coincide, and we are looking to determine whether the resulting one is stable. 

We employ the standard method for establishing linear stability: linearizing the system and  taking  $B = 2\sqrt{\frac{\csc(q_0)}{1- \cos(q_0)}}$ for some fixed $q_0$.

Computing the characteristic polynomial of the matrix of the linearized system at any relative equilibrium on the threshold curve gives us
\begin{equation*}
    x\left(x^2 + 2\csc^3(q_0)\right)\left(x^2 + 2(1 + 2\cos(q_0))\csc^3(q_0)\right)=0
\end{equation*}
with the solutions
\begin{equation*}
    \begin{split}
        x = 0,\  x = \pm\mathrm{\mathbf{i}}\sqrt{2}\csc^{\frac{3}{2}}(q_0),  \ x = \pm\sqrt{-2(2\cot(q_0)\csc^2(q_0)+\csc^3(q_0))}
    \end{split}
\end{equation*}

The zero is due to the Casimir being conserved, so the other four roots determine the linear stability of the equilibrium on the reduced space; in particular the sign of the expression under the root determines its linear stability.

It is greater than zero (rendering the equilibrium linearly unstable) when $\frac{2\pi}{3}<q_0<\pi$ and less than 0 with a linearly stable equilibrium when $0<q_0<\frac{2\pi}{3}$

We have demonstrated 
\begin{proposition} \label{prop:Type II stability}
For the values of $(q,B)$ on the threshold curve, the Type II equilibrium is
\begin{itemize}
    \item linearly stable if $(q,B)$ is to the left of\/ $\left(\frac{2\pi}{3}, \frac{4}{3}3^{1/4}\right)$,
    \item  linearly unstable if $(q,B)$ is to the right of\/ $\left(\frac{2\pi}{3}, \frac{4}{3}3^{1/4}\right)$.
\end{itemize}
\end{proposition}

\subsubsection{Stability of general relative equilibria of Type II}
Now we position ourselves in the region strictly above the threshold curve. 
Again, we linearize the system at the equilibrium point and look at the zeros of the characteristic polynomial. 
However, due to the complexity of the equations we have to employ numerical methods. 

\begin{figure}
\centering
\subfigure[$(m_2^-, m_3^-)$ solutions]{\includegraphics[scale = 0.8]{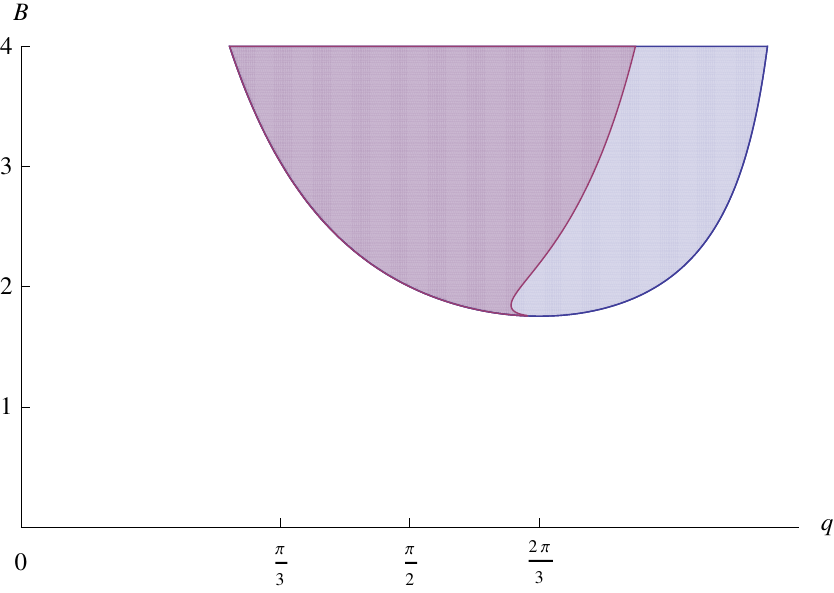}}\hfil
\subfigure[$(m_2^+, m_3^+)$ solutions]{\includegraphics[scale = 0.8]{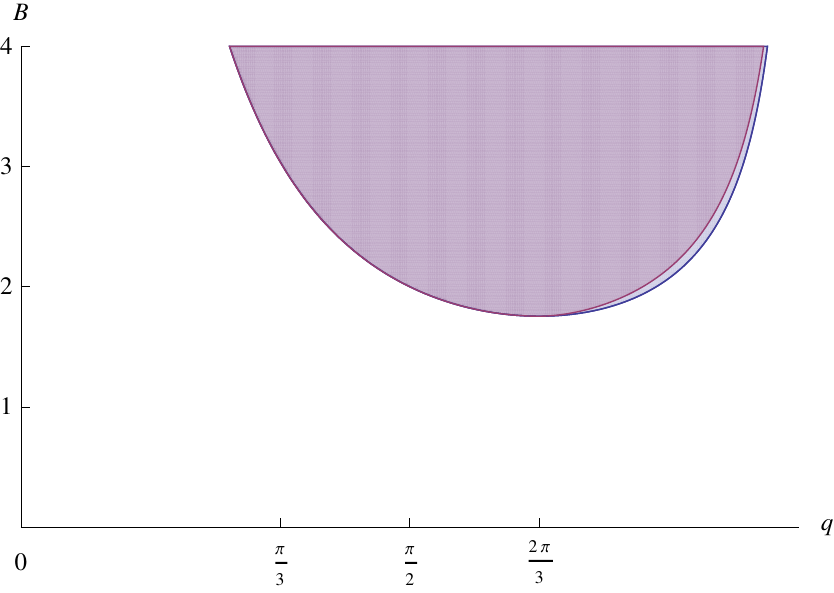}}
\caption{Stability of relative equilibria of Type II in terms of $B$ and $q$: in the dark purple regions, the corresponding relative equilibrium is linearly stable, while the blue colour denotes linear instability.}
\label{fig:Type II stability}
\end{figure}

We have performed the numerics for values of $B$ less than 100, and found that up to this value the properties are always as shown in Figure \ref{fig:Type II stability}.

As calculations demonstrate, the curves that separate the regions of stability from those of instability consist of degenerate relative equilibria; these are, in fact, the only degenerate relative equilibria of Type II. As will be discussed later, for a fixed value of $B$ the Casimir function assumes its minima and maxima on the relative equilibria lying on the curves where stability changes. 

\begin{figure}
\centering
\includegraphics[scale = 0.6]{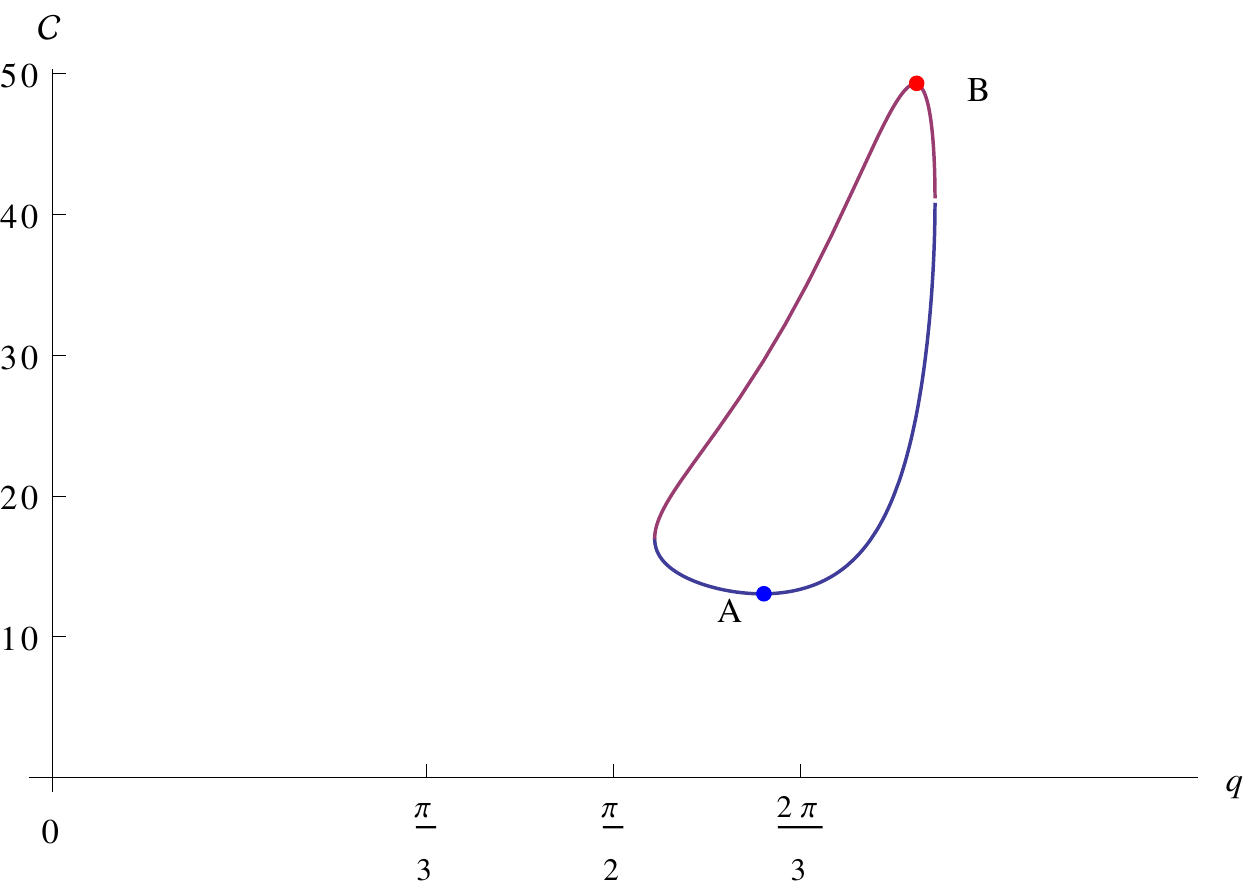}
\caption{The values of $\mathcal{C}$ for relative equilibria of Type II with a fixed $B = 1.9$. The blue (lower) part of the curve is given by $(m_2^-, m_3^-)$, the red (upper) one by $(m_2^+,m_3^+)$.}
\label{fig:Cas against B type casimir fixed b}
\end{figure}

In Figure \ref{fig:Cas against B type casimir fixed b}, the points A and B on the curve are the points where the Casimir is minimal and maximal respectively, on the family of relative equilibria for that fixed value of the magnetic field strength $B$.  Stable relative equilibria lie to the left of these points and unstable ones to the right.  These two points represent saddle-node bifurcations of the relative equilibria of Type II as the value of the Casimir is varied. 

 For both $(m_2^+,m_3^+) $ and $(m_2^-,m_3^-) $ the curves dividing the stability region from that of instability separate from the threshold curve at the point $(q,B) = \left(\frac{2\pi}{3}, \frac{4}{3}3^{1/4}\right)$

\subsubsection{Energy-Casimir map revisited}

\begin{figure}\centering
\subfigure[Bifurcation diagram for relative equilibria of Type I.]{
    \includegraphics[scale  = 0.45]{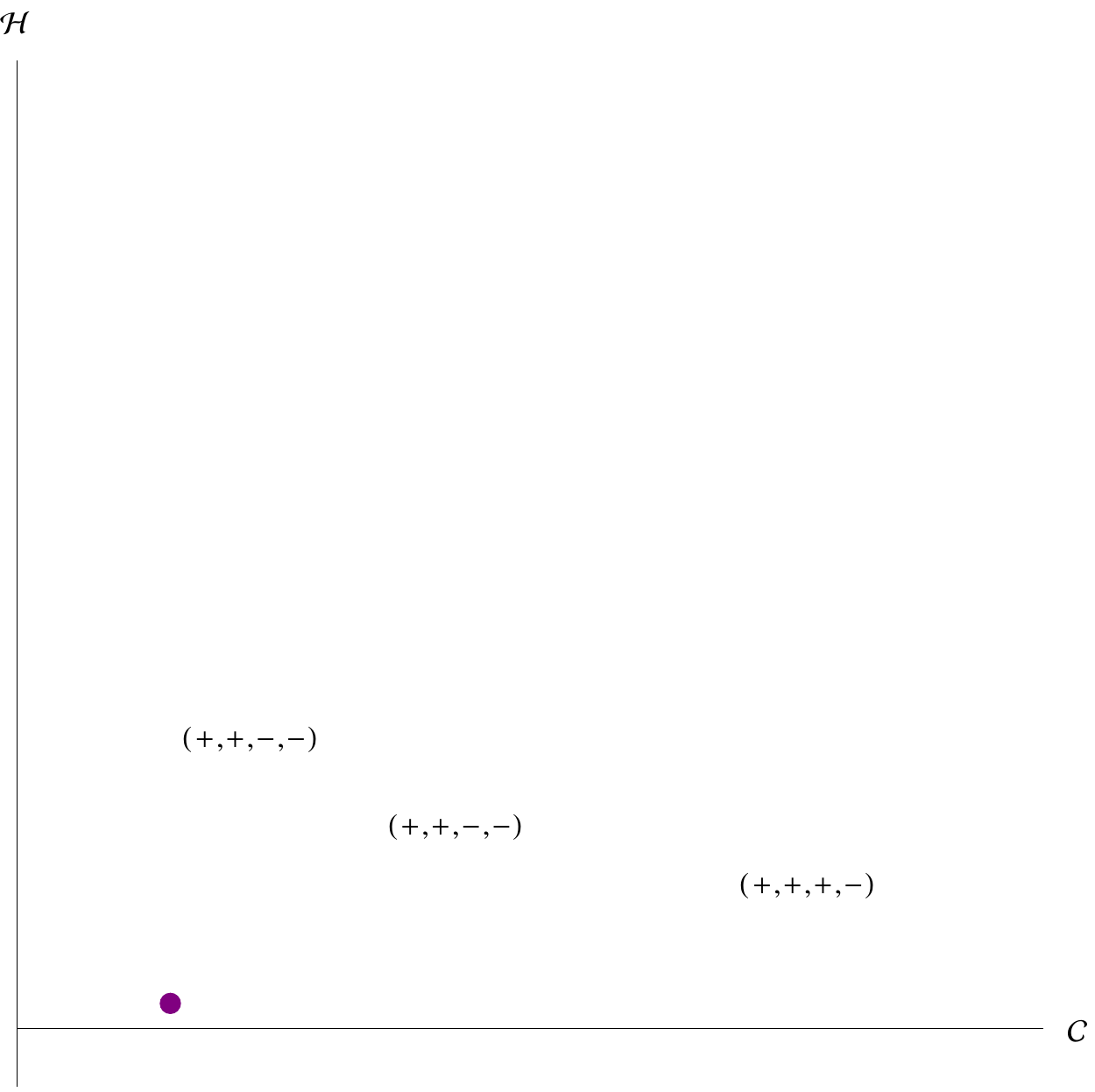}
} \hfil
\subfigure[Schematic  bifurcation diagram for relative equilibria of Type II: the blue portion of the curve is for $(m_2^-,m_3^-)$ and the red for $(m_2^+,m_3^+)$.]{
    \includegraphics[scale  = 0.45]{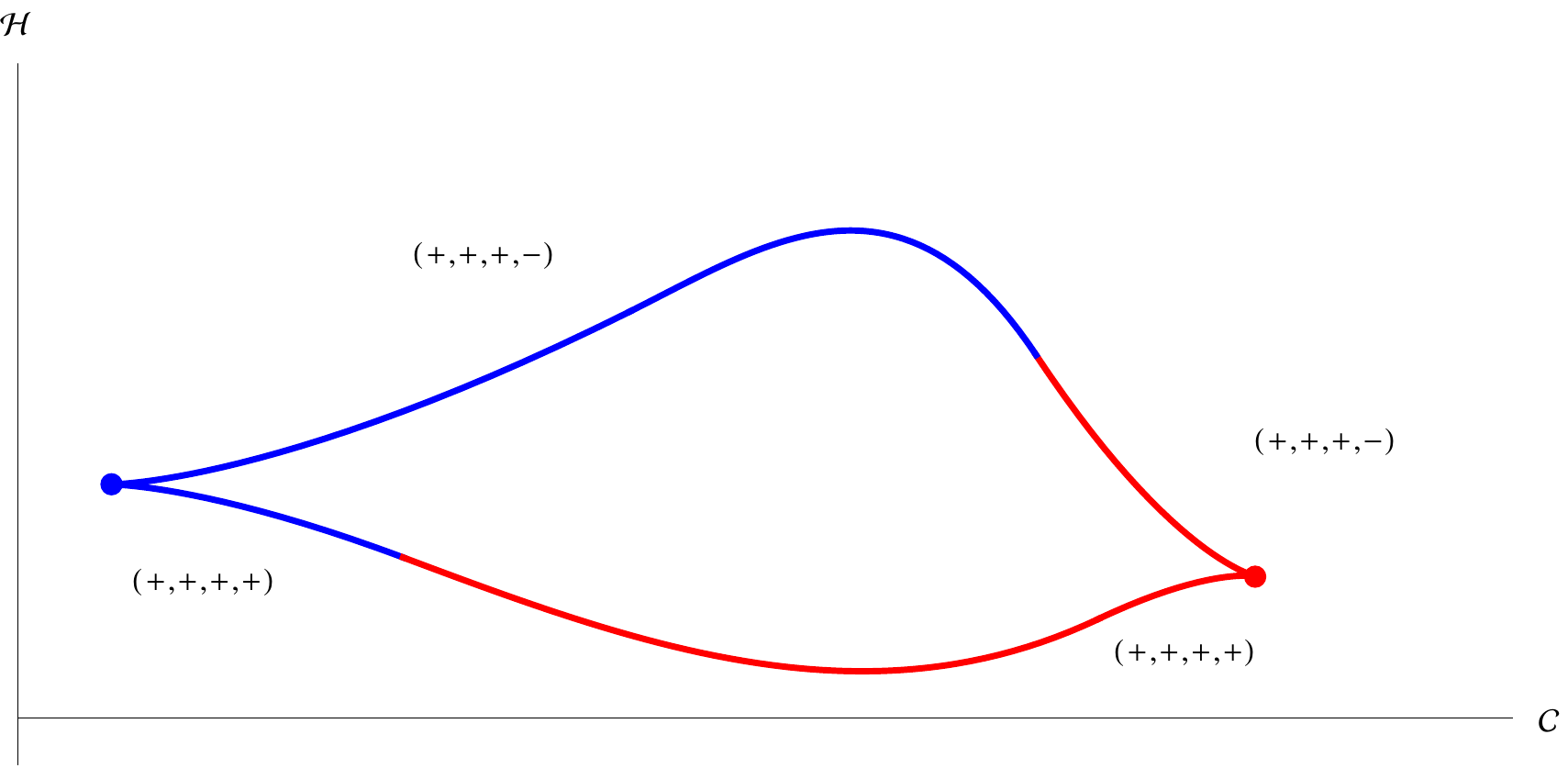}
}    \caption{Energy-Casimir bifurcation diagrams with cusp points and signatures of the Hessian.}
    \label{fig:schematic energy-momentum}
\end{figure}

With the newly acquired information about the stability, we cast another look at the Energy-Casimir map.  Figure \ref{fig:schematic energy-momentum} (a) and (b) shows the set of singular values of this map, which are the images of the set of relative equilibria. The figure in (b) is schematic, as the two branches are very close in reality. 
The cusps on the curves, emphasized by dots, are the configurations at which the transition between stability and instability occurs. They represent saddle-node bifurcations when using the Casimir as a parameter. The cusps in (b) correspond to the points marked A and B in Figure \ref{fig:Cas against B type casimir fixed b}. 

The topmost curve in Figure \ref{fig:schematic energy-momentum}(a) corresponds to the values of $q> \pi/2$, the lower half of the bottom curve  to the values of $q$ less than the root (as solved for $q$) of $B = \sqrt{\frac{\cos^3(q)(2 + \cos(q))}{2\sin^3\left(q\right)\sin^2\left(\frac{q}{2}\right)}}$ for a fixed $B$, and the upper half  to the rest of the interval between said root and $\pi/2$. (The Figure is very similar to \cite[Fig.\ 9]{borisov2018reduction}, of which it is a continuation.)

We have simplified the form of the bifurcation diagram in Figure  \ref{fig:schematic energy-momentum}(b), but the essential features remain: the two solutions, one of which is linearly stable and the other linearly unstable, merge together at the two cusps in saddle-node bifurcations. 

Figure \ref{fig:schematic energy-momentum} has the signatures of the Hessian of $\mathcal{H}$, as restricted to the level sets of $\mathcal{C}$ next to each part of the bifurcation diagram. Since the eigenvalues of the Hessian depend smoothly on $B$ and $q$ and the only points where the Hessian matrix has a zero eigenvalue are the cusp points (the only points where the matrix of the linearized system has a zero eigenvalue), it is sufficient to calculate the signature of the Hessian on each part of the diagram for a single value of $B$ and $q$.  By continuity, the signatures will remain the same throughout the changes in $B$ and $q$.  

It follows that the linearly stable (relative) equilibria in Proposition \ref{prop:Type II stability} are in fact nonlinearly stable.

\begin{rmk} \label{rmk:stability v. axis}
It is curious that for the Type I relative equilibria, the unstable configurations occur when the particles are closer together, but for the Type II relative equilibria, the unstable ones occur when they are further apart. However, in both cases, the unstable ones occur when the particles are further from the axis of rotation (see Figure \ref{fig:acute and obtuse RE}). 
\end{rmk}

\subsection{The bifurcation diagram}
So far, we have described the existence of relative equilibria in terms of $B$ and $q$. This is a reasonable approach for presenting the results, but carries no physical meaning in terms of bifurcations of dynamical systems.  Indeed, for each fixed value of $B$, there is a 1-parameter family of reduced systems parametrized naturally by the Casimir. We therefore have two parameters: the Casimir (an `internal' parameter) and the magnetic field strength (an `external' parameter).

For relative equilibria of Type I, parametrized by $B$ and $q$, the explicit expression for the Casimir is
$$\frac{\cos(q/2)}{\sin^3(q/2)}\left(1+B^2\sin(q)\tan^2(q)\right),$$
which for all values of $B$ tends to $+\infty$ if $q\to 0$ and to $0$ if $q\to \pi$, spanning all the values in between. Therefore, the region spanned by all possible Casimir values as $B$ varies is the entirety of the first quadrant in the $(B,\mathcal{C})$-plane.

\begin{figure}
\centering
\subfigure[$(m_2^-,m_3^-)$]{\includegraphics[scale = 0.85]{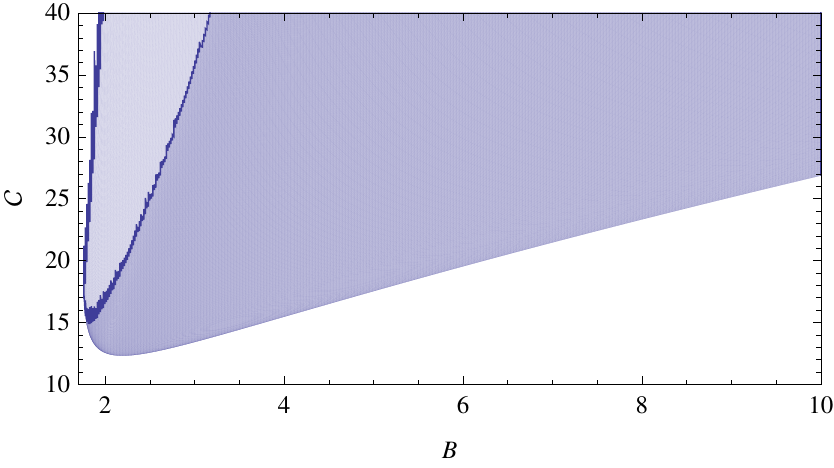}}
\subfigure[$(m_2^+,m_3^+)$]{\includegraphics[scale = 0.85]{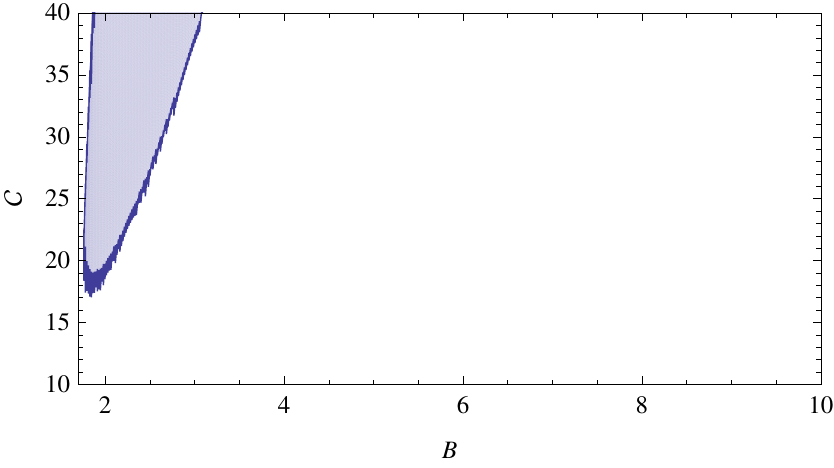}}
\caption{Possible values of the Casimir for the two classes of relative equilibria of Type II, as a function of $B$.}
\label{fig:Cas against B type II both cases}
\end{figure}

The situation is more complicated when we pose the same problem for relative equilibria of Type II: once again, we have to resort to numerics due to the complexity of the computations. The image of the set of relative equilibria for $(m_2^-, m_3^-)$ is depicted in Figure \ref{fig:Cas against B type II both cases}(a).
The darker blue region on the diagram denotes the locus of the points in the $(B,\mathcal{C})$ plane that are images of two relative equilibria.

For $(m_2^+,m_3^+)$ (Type II) the possible values of the Casimir lie inside the set depicted in Figure \ref{fig:Cas against B type II both cases}(b). 

As can be noted, the region in the second diagram fits in the lighter area of the first one, which is clear since every point strictly inside the union of the two sets corresponds to two values of $q$ and, therefore, to two relative equilibria. 

The image in the $(B,\mathcal{C})$-plane of the threshold curve is the transition between the two regions in Figure \ref{fig:Cas against B type II both cases} and is shown as a dark blue curve in both.

Since the values of the Casimir for the relative equilibria of Type II, as plotted against $q$ (see Figure \ref{fig:Cas against B type casimir fixed b})  form a closed curve, the set of possible values of $\mathcal{C}$ on the set of relative equilibria is bounded for every value of $B$. Saddle-node bifurcations arise at the extreme points of $\mathcal{C}$ on the curve.  

These relative equilibria are degenerate; they coincide with the set of cusp points  for Type II relative equilibria in Figures \ref{fig:Ham Cas map image} and \ref{fig:schematic energy-momentum} and, as the only  degenerate relative equilibria of Type II, with the curves in Figure \ref{fig:Ham Cas map image} that separate the regions of stability and instability. Therefore, the bifurcation curve is the image of said curves on the $(B, \mathcal{C})$ plane.


Indeed, since the relative equlibria on the bifurcation curve are degenerate, the only point where it can meet the threshold curve is the only degenerate relative equilibrium on the threshold curve: $B = \frac{4}{3}3^{1/4},\; \mathcal{C} = {20}/(3\sqrt{3})$, the left-most point on the image of the threshold curve in Figure \ref{fig:Cas against B type II both cases}.

From the discussion above it can be easily seen that for every fixed value of $B$  and $\mathcal{C}$ for which there are 2 relative equilibria of Type II, one of these is (linearly)  stable the other unstable.

\section{Opposite charges} \label{sec:opposite charges}

After describing in some detail the relative equilibria with equal charges, the next logical step is to investigate the case with opposite charges, thereby replacing the repelling potential by an attracting one. The following observation is based on \cite[Lemma 2.4]{garcia2020attracting}, which there is for a Lagrangian system and here includes the magnetic field.

On the configuration space $\mathcal{Q}=S^2\times S^2\setminus\Delta$ (see Sec. \ref{sec:Motion of two particles}), the involution $(\mathbf{q}_1,\mathbf{q}_2) \mapsto(\mathbf{q}_1,-\mathbf{q}_2)$ is antisymplectic on the second component. If this is combined with a change of sign of charge $e_2\mapsto -e_2$ of the second particle, then the symplectic form in \eqref{omega} is unchanged (we are not changing the sign of the magnetic field). 

Write $V_1(\mathbf{q}_1,\mathbf{q}_2)$ for the potential energy obtained from $V(\mathbf{q}_1,\mathbf{q}_2)$ after changing $e_2$ to $-e_2$. Then if $V(\mathbf{q}_1,\mathbf{q}_2)=V_1(\mathbf{q}_1,-\mathbf{q}_2)$ then the involution transforms the Hamiltonian to itself.  Under this condition, the involution will map the Hamiltonian system with potential energy $V$ to the one with potential energy $V_1$. 

A case in point is $V(\mathbf{q}_1,\mathbf{q}_2)=e_1e_2\cot(q)$ where $q$ is the geodesic distance between $\mathbf{q}_1$ and $\mathbf{q}_2$ on the sphere. For then $V_1(\mathbf{q}_1,\mathbf{q}_2) = e_1(-e_2)\cot(q)$, and $V(\mathbf{q}_1,-\mathbf{q}_2) = e_1e_2\cot(\pi-q)=-e_1e_2\cot(q)$. 

The effect of this involution on the reduced coordinates is
$$(m_1,m_2,m_3,q,p) \longmapsto (-m_1,-m_2,m_3,\pi-q,-p)$$
(notice that $m_3$ is unchanged). 
The Casimir is invariant, and so is the Hamiltonian provided the potential changes as discussed above. Each relative equilibrium, as well as its stability properties, for charges $(e_1,e_2)$ is therefore mapped to a relative equilibrium for charges $(e_1,-e_2)$, together with its stability properties, by this involution. 

Now consider the specific potential $V(q)=e_1e_2\cot q$ (repelling for like charges, attracting for charges of opposite sign).  All the conclusions about the relative equilibria found in Section 5 carry over here, up to reflection of all the graphs with respect to the line $q = \frac{\pi}{2}$. For example, the threshold curve becomes $B= 2\sqrt{\frac{\csc(q)}{1+\cos(q)}}$. In the same way, the relative equilibria are divided into two types, I and II, depending on domains of existence. We refer to them accordingly, depending on the one they coincide with via the involution described above.

\begin{figure}[t] 
\centering

\subfigure[Obtuse and acute relative equilibria of Type I.]{%
    \begin{tikzpicture}[scale=1.5]  
        \draw [thick] (0,0) circle (1cm);
            \draw [-latex,dashed] (0,-1) -- (0,1.5);
            \draw [-latex] (0.2,1.25) arc (-30:250: 2mm and 0.5mm);
            \draw [-latex] (0.2,1.25) arc (-30:250: 2mm and 0.5mm);
            \draw (-0.3,1.3) node {$\boldsymbol\omega$};
            \draw [thin] (0,0) -- (0.819,-0.573);
            \draw (0,0.3) arc (90:105:3mm);
            \filldraw [violet] (0.819,-0.573) circle (2pt);
            \draw [thin] (0,0) -- (-0.2588, 0.9659);
            \draw (0,0.2) arc (90:-37:2mm);
            \draw (0,0.23) arc (90:-37:2.3mm);
            \filldraw [violet] (-0.2588, 0.9659) circle (2pt);
     \end{tikzpicture}
         \hskip 6mm
    \begin{tikzpicture}[scale=1.5] 
        \draw [thick] (0,0) circle (1cm);
            \draw [-latex,dashed] (0,-1) -- (0,1.5);
            \draw [-latex] (0.2,1.25) arc (-30:250: 2mm and 0.5mm);
            \draw (-0.3,1.3) node {$\boldsymbol\omega$};
            \draw [thin] (0,0) -- (-0.642,0.776);
            \draw (0,-0.2) arc (-90:-230:2mm);
            \filldraw [violet] (-0.642,0.776) circle (2pt);
            \draw [thin] (0,0) -- (0.1736, 0.984);
            \draw (0,-0.2) arc (-90:80:2mm);
               \draw (0,-0.25) arc (-90:80:2.5mm);
            \filldraw [violet] (0.1736, 0.984) circle (2pt);
 \end{tikzpicture}
 } 
 \hskip15mm
\subfigure[Obtuse and acute relative equilibria of Type II]{%
{\begin{tikzpicture}[scale=1.5]  
        \draw [thick] (0,0) circle (1cm);
            \draw [-latex] (0.2,1.5) arc (-30:250: 2mm and 0.5mm);
            \draw [-latex,dashed] (0,-1) -- (0,1.4);
              \draw [-latex,dashed] (0,-1) -- (0,1.8);
            \draw [-latex] (0.2,1.15) arc (-30:250: 2mm and 0.5mm);
            \draw (-0.35,1.15) node {$\boldsymbol\omega_1$};
            \draw (-0.35,1.5) node {$\boldsymbol\omega_2$};
            \draw [thin] (0,0) -- ( 0.4226,0.9063);
            \draw (0,0.2) arc (90:65:2mm);
            \draw [thin] (0,0) -- (0.4226,-0.9063);
            \draw (0,-0.2) arc (-90:-65:2mm);
             \filldraw [violet] (0.4226,0.9063) circle (2 pt);
            \filldraw [violet] (0.4226,-0.9063) circle (2 pt);
 \end{tikzpicture}}
         \hskip 6mm
\begin{tikzpicture}[scale=1.5]  
        \draw [thick] (0,0) circle (1cm);
            \draw [-latex,dashed] (0,-1) -- (0,1.4);
            \draw [-latex,dashed] (0,-1) -- (0,1.8);
            \draw [-latex] (0.2,1.15) arc (-30:250: 2mm and 0.5mm);
             \draw [-latex] (0.2,1.5) arc (-30:250: 2mm and 0.5mm);
            \draw (-0.35,1.15) node {$\boldsymbol\omega_1$};
            \draw (-0.35,1.5) node {$\boldsymbol\omega_2$};
            \draw [thin] (0,0) -- ( 0.939, -0.342);
            \draw (0,-0.2) arc (-90:-20:2mm);
            \filldraw [violet] (0.939, -0.342) circle (2 pt);
            \draw [thin] (0,0) -- (0.939, 0.342);
            \draw (0,0.2) arc (90:20:2mm);
            \filldraw [violet] (0.939, 0.342) circle (2 pt);
     \end{tikzpicture}}
     
 \caption{The two types of relative equilibrium for equal masses and opposite charges, with $V(q) = -\cot(q)$.}
 \label{fig:RE opposite charges}
 \end{figure}
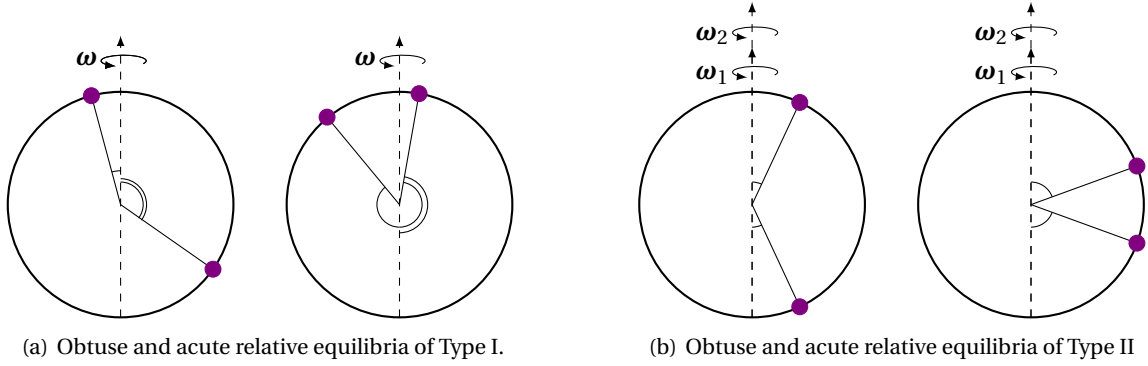

The geometric differences with the case of identical particles is illustrated in Figure \ref{fig:RE opposite charges}. In particular, the new Type I relative equilibria  have the axis between the particles, and the configuration continues to be asymmetric. (Compare with the analogous transformation in \cite{garcia2020attracting}.)

Type II relative equilibria are no longer isosceles configurations in the sense we have used before; however, they retain a symmetry, with the axis of rotation lying to the side of the particle pair. As previously, the same geometric arrangement can be occupied by relative equilibria with two distinct rates of rotation (about the same axis), and hence two different energy levels. 

The plots of the regions of stability and instability shown in Figures \ref{fig:Type I stability} and \ref{fig:Type II stability} remain the same, except for a reflection in the line $q=\pi/2$. 

In particular (cf.\ Remark \ref{rmk:stability v. axis}) relative equilibria with configurations that are closer to the axis of rotation are now more likely to be stable.

\appendix

\section{A limiting argument}\setcounter{equation}{0} \def\theequation{A.\arabic{equation}}
Here we elaborate on the behaviour of the solutions in \eqref{eq:idensolsa} when $q\to\frac{\pi}{2}, \ B\to 0$. 

In the case of equal masses and a repelling potential, right angle equilibria exist for the two-body problem on a sphere \cite{garcia2020attracting}. However, they are not defined for the case when $B\ne 0 $ and is less than the minimum value of the threshold curve. On the other hand, seeing that the equal masses with a repelling potential  gravitational two-body problem is a limiting case with $B\to 0$ for two identical particles, these equilibria should arise from the ones that we have described.

When $B = 0$, the system \eqref{two id part equat} is reduced to one equation
\begin{equation}
    \label{equations when B = 0}
    m_2m_3 = 1,
\end{equation}
giving a family of right angle equilibria in accordance with \cite{garcia2020attracting}. We have already mentioned that setting $B$ equal to 0 and taking the Taylor series at $q =\frac{\pi}{2}$ gives finite limits in the cases of the solutions \eqref{eq:idensolsa}. Indeed, we get 
\begin{align}
& m_2^{\pm} =\mp1\pm \frac{3}{2} \left(q-\frac{\pi }{2}\right) +  \bar{O}\left(q - \frac{\pi}{2}\right)^2\\
 &m_3^{\pm} = \mp1 \pm \frac{1}{2} \left(q-\frac{\pi }{2}\right) + \bar{O}\left(q - \frac{\pi}{2}\right)^2
\end{align}
However, setting $q=\frac{\pi}{2}$ and \textbf{then} taking $B=0$ results in an indefinite expression. To explain this, let us consider the expression $m_2^{\pm}m_3^{\pm}$. 

\begin{equation}
\begin{split}
   m_2^{\pm}m_3^{\pm} = &\pm \frac{1}{4} \cot \left(\frac{q}{2}\right) \left(\sqrt{B^2 \sin ^2(q) \tan ^2(q)+4 \cot \left(\frac{q}{2}\right)}-B \sin (q) \tan (q)\right)* \\ &
   *\left(\sqrt{B^2 \sin ^2(q) \tan^2(q)+4 \cot \left(\frac{q}{2}\right)}+B (\cos (q)+\sec (q)-2)\right)
  \end{split}
\end{equation}
When $B\to0$, 
\begin{equation}
    m_2^{\pm}m_3^{\pm}\to \cot^2\left(\frac{q}{2}\right) ,
\end{equation}
but it can easily be seen that $m_2^{\pm}m_3^{\pm}$ does not converge uniformly to  $ \cot^2\left(\frac{q}{2}\right)$ in the neighbourhood of $q = \frac{\pi}{2}$ with $B\to0$ (see Part III, Chapter XVI of \cite{arkhipov1999lectures} for definitions). Thus, the order of the limits can't be changed.

However, we can take $B$ as a function of $q$, demand that $B\left(\frac{\pi}{2}\right) = 0$ and see whether a limit exists when $q \to \frac{\pi}{2}$. 

Calculation of Taylor series of $m_2^{\pm}$ and $m_3^{\pm}$ shows  that only the linear approximation of $B(q)$ (i.e. $B'\left(\frac{\pi}{2}\right)$) plays a role in the behaviour at $q =\frac{\pi}{2}$. In fact, if $B'\left(\frac{\pi}{2}\right) = a$, we have
\begin{equation}
\begin{split}
   & m_2^{\pm}= \frac{1}{2} \left(a\mp\sqrt{a^2+4}\right) + \bar{O}\left(q - \frac{\pi}{2}\right)\\
   & m_3^{\pm} = \frac{1}{2} \left(\mp\sqrt{a^2+4}-a\right) +  \bar{O}\left(q - \frac{\pi}{2}\right),
    \end{split}
\end{equation}
resulting in
\begin{equation}
    m_2^{\pm}m_3^{\pm} = 1  + \bar{O}\left(q - \frac{\pi}{2}\right),
\end{equation}
with different directions of approaching the point $(q,B) = (\frac{\pi}{2}, 0)$ giving us different instances of right angle relative equilibria for the two body problem.

Note that the set of right-angle relative equilibria is "wrapped" into one point on $(q,B)$-plane. It is precisely the non-uniform convergence of the solutions that allows us to approximate the whole family of right-angle relative equilibria rather than just one: indeed, for every small value of $B$ we can find a stalk of functions $B(q)$ such that respective relative equilibria approximate the given right-angle one.

\subsection*{But what happens to  right-angle relative equilibria?}
As was described above, right-angled equilibria do not exist until the value of $B$ reaches a certain threshold. 

Suppose now that the particles are in a right angle equilibrium state, with $B=0$, and we "switch on" the magnetic field. What happens to the particle motion?

When the newly appeared magnetic field is weak ($B<\frac{4}{3}3^{1/4}$) for a right angle relative equilibrium, we land in an initial state with a non-zero $B$ and $q = \frac{\pi}{2}$, which can not be a relative equilibrium, and neither it can turn into one with the passage of time, since our system is deterministic. 

For a stronger magnetic field,  we theoretically might achieve a Type II relative equilibrium. As mentioned above, from the equations \eqref{two id part equat} with $B=0$ and $q = \frac{\pi}{2}$, the conditions for relative equilibria are $p=0,\ m_1 = 0,\ m_2m_3 = 1$. On the other hand, the product of $m_2$ and $m_3$ for any Type II relative equilibrium is always negative.

Thus, for any change in the strength of the magnetic field, the right angle relative equilibria do not persist.

\paragraph{Remark}
It would be interesting to analyze this 2-body problem from a control theoretic perspective, where $B$ is the control parameter.

\setlength{\parindent}{0pt}\small

\hrulefill

\bigskip 

{N. Balabanova \& J.~Montaldi} \\
Department of Mathematics, \\
University of Manchester \\
Manchester M13 9PL, UK\\
\texttt{nataliya.balabanova@manchester.ac.uk}\\
\texttt{j.montaldi@manchester.ac.uk}

\end{document}